\documentclass{aa}  

\usepackage{graphicx}
\usepackage{txfonts}
\usepackage{xcolor}
\usepackage{graphicx}
\usepackage{subcaption}
\usepackage{titlesec}
\usepackage{multirow}
\usepackage{booktabs}   
\usepackage{xcolor}
\usepackage{array}
\usepackage{orcidlink}
\usepackage{hyperref}

\begin{document} 

\title{Energy-resolved pulse profiles of Vela X-1: cross-calibrating XMM-Newton and NuSTAR to trace spectral features}

  
\author{
Dimitrios K. Maniadakis\orcidlink{0009-0008-1148-2320} \thanks{Email: \href{mailto:dim.maniad@gmail.com}{dim.maniad@gmail.com}} \inst{1,2}
\and
Antonino D'A\`i \inst{1}\orcidlink{0000-0002-5042-1036}
\and
Camille M. Diez \inst{3}\orcidlink{0000-0001-6520-4600}
\and
Giancarlo Cusumano \inst{1}\orcidlink{0000-0002-8151-1990}
\and
Elena Ambrosi \inst{1}\orcidlink{0000-0002-9731-8300}
\and
Carlo Ferrigno \inst{4,5}\orcidlink{0000-0003-1429-1059}
\and
Ekaterina Sokolova-Lapa \inst{6}\orcidlink{0000-0001-7948-0470}
\and
Matteo Lucchini \inst{7}\orcidlink{0000-0002-2235-3347}
\and
Peter Kretschmar \inst{3}\orcidlink{0000-0001-9840-2048}
\and
Alessio Anitra \inst{2}\orcidlink{0000-0002-2701-2998}
\and
Christian Malacaria \inst{8}\orcidlink{0000-0002-0380-0041}
\and
Gabriele A. Matzeu \inst{9}\orcidlink{0000-0003-1994-5322}
\and
Ciro Pinto \inst{1}\orcidlink{0000-0003-2532-7379}
\and
J\"orn Wilms \inst{6}\orcidlink{0000-0003-2065-5410}
\and
Felix F\"urst \inst{3}\orcidlink{0000-0003-0388-0560}
}

\institute{
INAF - IASF Palermo, via Ugo La Malfa 153, 90146 Palermo, Italy
\and
Dipartimento di Fisica e Chimica Emilio Segr\`e, Universit\`a degli Studi di Palermo, Via Archirafi 36, 90123 Palermo, Italy
\and
European Space Agency (ESA), European Space Astronomy Centre (ESAC), Camino Bajo del Castillo s/n, 28692 Villanueva de la Ca\~nada, Madrid, Spain
\and
Department of Astronomy, University of Geneva, Chemin d'\'Ecogia 16, CH-1290 Versoix, Switzerland
\and
INAF, Osservatorio Astronomico di Brera, Via E. Bianchi 46, I-23807 Merate, Italy
\and
Dr. Karl Remeis-Observatory and Erlangen Centre for Astroparticle Physics, Friedrich-Alexander-Universit\"at Erlangen-N\"urnberg, Sternwartstr. 7, 96049 Bamberg, Germany
\and
Anton Pannekoek Institute for Astronomy, University of Amsterdam, Science Park 904, 1098 XH Amsterdam, the Netherlands
\and
INAF Osservatorio Astronomico di Roma, via di Frascati 33, 00078 Monteporzio Catone, Roma, Italy
\and
Quasar Science Resources SL for ESA, European Space Astronomy Centre (ESAC), Science Operations Department, 28692 Villanueva de la Ca\~nada, Madrid, Spain
}

   \date{}

 
  \abstract
   {
   Pulse profiles probe the emission geometry of accreting X-ray pulsars, but their observed shapes may depend on 
   instrumental response and observational setup. The pulsed fraction spectrum provides a compact spectro-timing 
   observable that can both trace localized spectral features and serve as a quantitative cross-calibration diagnostic.} 
   {
    We assess the consistency of energy-resolved pulse profiles obtained with simultaneous
    \textit{XMM-Newton}/EPIC-pn and \textit{NuSTAR}/FPM observations of Vela~X-1, and investigate 
    the broadband pulsed fraction spectrum as a diagnostic of spectral features from 1 to 70\,keV.
   }
   {
    We construct energy–phase matrices for both instruments and derive 
    pulsed fraction spectra after carefully accounting for instrumental and observational effects. We quantify the residual systematics 
    in the overlapping 3–10\,keV band. We, then, model the broadband pulsed fraction spectra phenomenologically and search for timing signatures of spectral features.
   }
   {
    After correcting for instrumental effects, the pulsed fraction spectra derived strictly over the common exposure intervals of the 
    two instruments agree within 5\% in their overlapping 3--10\,keV range. Remaining discrepancies larger than $5\%$ are confined
    to the iron-line region and can be attributed to the different energy resolutions of the two instruments. The broadband pulsed fraction
    spectrum reveals significant localized features corresponding to known emission lines in the soft band and to the fundamental and harmonic 
    cyclotron resonant scattering features at $\sim$25 and $\sim$55\,keV. An orbital-phase-resolved modeling of the pulsed fraction spectrum 
    of EPIC-pn shows that the soft band features strongly depends on the value of the equivalent absorption column, with emission line signatures becoming progressively 
    suppressed during highly absorbed intervals. 
   }
   { The pulsed fraction spectrum serves both as a quantitative cross-calibration diagnostic and as a powerful spectro-timing diagnostic. Its modeling provides independent constraints on spectral features, complementing traditional phase-averaged spectroscopy. 
   }
\titlerunning{Energy-resolved pulse profiles of Vela~X-1}
\authorrunning{Maniadakis et al.}
   \maketitle
%
\section{Introduction}
\label{intro}


High-mass X-ray binaries (HMXBs) are binary systems in which a neutron star (NS) accretes material from an OB supergiant or Be-type donor star, converting gravitational potential energy into X-ray emission. The observed radiation originates from a combination of thermal and non-thermal processes, including bremsstrahlung, cyclotron emission, and inverse Compton scattering \citep{caballero2012x, Mushtukov2022a}. Most NS HMXBs appear as X-ray pulsars: their strong magnetic fields channel accreting plasma onto the magnetic poles, producing highly anisotropic emission modulated at the NS spin period \citep{bildsten1997observations}. These systems are unique laboratories for studying accretion under extreme conditions and the complex interaction between X-ray radiation and the surrounding environment \citep{Martinez-Nunez2017, Kretschmar2021}.

X-ray pulse profiles, defined as the folded light curve of the spin-modulated emission, carry information about the anisotropy of the emergent radiation field and the geometry-dependent visibility of the emission regions. They are influenced by reprocessing in the circumstellar material and by the beaming patterns associated with the underlying emission mechanisms \citep{Meszaros1985, Becker2012, sokolova2023radiative}.  However, this information is inevitably filtered by the characteristics of the instrument used to measure the pulse profiles. Instrumental effects can alter the observed shape through factors such as energy and temporal resolution, effective-area variations, photon redistribution, flux calibration accuracy, and the separation of intrinsic source emission from instrumental and astrophysical background contributions. Observational conditions and acquisition modalities can further influence the reconstructed pulse shape, including exposure time, source position within the detector, and the presence of stray light or contaminating signals \citep{ducci2023modeling}. A clear methodology to quantify the instrumental dependence of pulse profile shapes is still lacking.

In this context, the pulsed fraction (PF) provides a compact and quantitative description of the pulse modulation. Although several operative definitions exist (see \citealt{Ferrigno2023}), its meaning remains related to the strength of the modulated emission relative to the total flux. Its energy dependence - the PF spectrum - encodes how the pulse amplitude varies across the X-ray band. As such, it offers a natural framework to assess instrumental effects in an energy-resolved manner. Recently, \citet{Ferrigno2023} introduced a dedicated pipeline for the analysis of \textit{NuSTAR} pulse profiles and demonstrated the diagnostic potential of the PF spectrum as a spectro-timing tool. Even phenomenological modeling of the PF spectrum reveals localized variations at energies consistent with known spectral features, such as cyclotron resonant scattering features (CRSFs) and the Fe~\textsc{K}$\alpha$ emission line \citep[for current applications, see ][]{D'Ai2025, maniadakis2025, Ambrosi2026}. 

We therefore extend the application of the PF spectrum to \textit{XMM-Newton} \citep{jansen2001xmm}, focusing on the EPIC-pn instrument because of its large effective area and good energy resolution. We concentrate on EPIC-pn data acquired in Timing Mode, which provides the highest available timing resolution ($\sim30\,\mu$s) and the highest effective area. This extension enables us to probe the soft X-ray band and to directly compare pulse modulation properties with \textit{NuSTAR} in their overlapping energy range, using the PF spectrum both as a tracer of spectral features and as a quantitative cross-calibration diagnostic between the two instruments.

To apply this methodology in a well-defined physical context, we consider Vela~X-1, a prototypical wind-accreting X-ray pulsar with a pulse period of ${\sim}283$\,s and a neutron-star mass of ${\sim}1.8\,M_\odot$ \citep{Rawls2011,Falanga2015,Kretschmar2021}. The source exhibits strong flux and spectral variability driven by a highly structured stellar wind. The absorption column density varies systematically with orbital phase, decreasing after eclipse to a minimum at $\phi\sim0.2$--0.3 and rising sharply around mid-orbit \citep{Sato1986,Haberl1990,Lewis1992}, reflecting the NS passage through different wind regions. These modulations are commonly attributed to large-scale structures such as accretion and photo-ionization wakes \citep{Fryxell1987,Blondin1990,Fransson1980} or a stable tidal stream \citep{Blondin1991,Blondin1994}, superimposed on an intrinsically clumpy wind that causes rapid absorption variability, with $N_\mathrm{H}$ spanning from a few $\times 10^{22}\,\mathrm{cm}^{-2}$ up to $\sim 10^{24}\,\mathrm{cm}^{-2}$  \citep{Lucy1980,Oskinova2007,Malacaria2016,MartinezNunez2017,Grinberg2017}. X-ray spectroscopy has revealed strong Fe~K$\alpha$ fluorescence and rich emission-line spectra \citep{Becker1978,Ohashi1984,Nagase1994,Sako1999,Schulz2002}, consistent with a two-phase medium whose properties depend on orbital phase and absorption state \citep{Watanabe2006,Goldstein2004,Grinberg2017,Lomaeva2020,Amato2021}. Despite extensive observational and theoretical work \citep{Blondin1994,Manousakis2015,Kretschmar2021}, the dominant nature of the absorbing structures remains debated. 

\begin{figure*}
    \centering
    \includegraphics[width=1.0\textwidth]{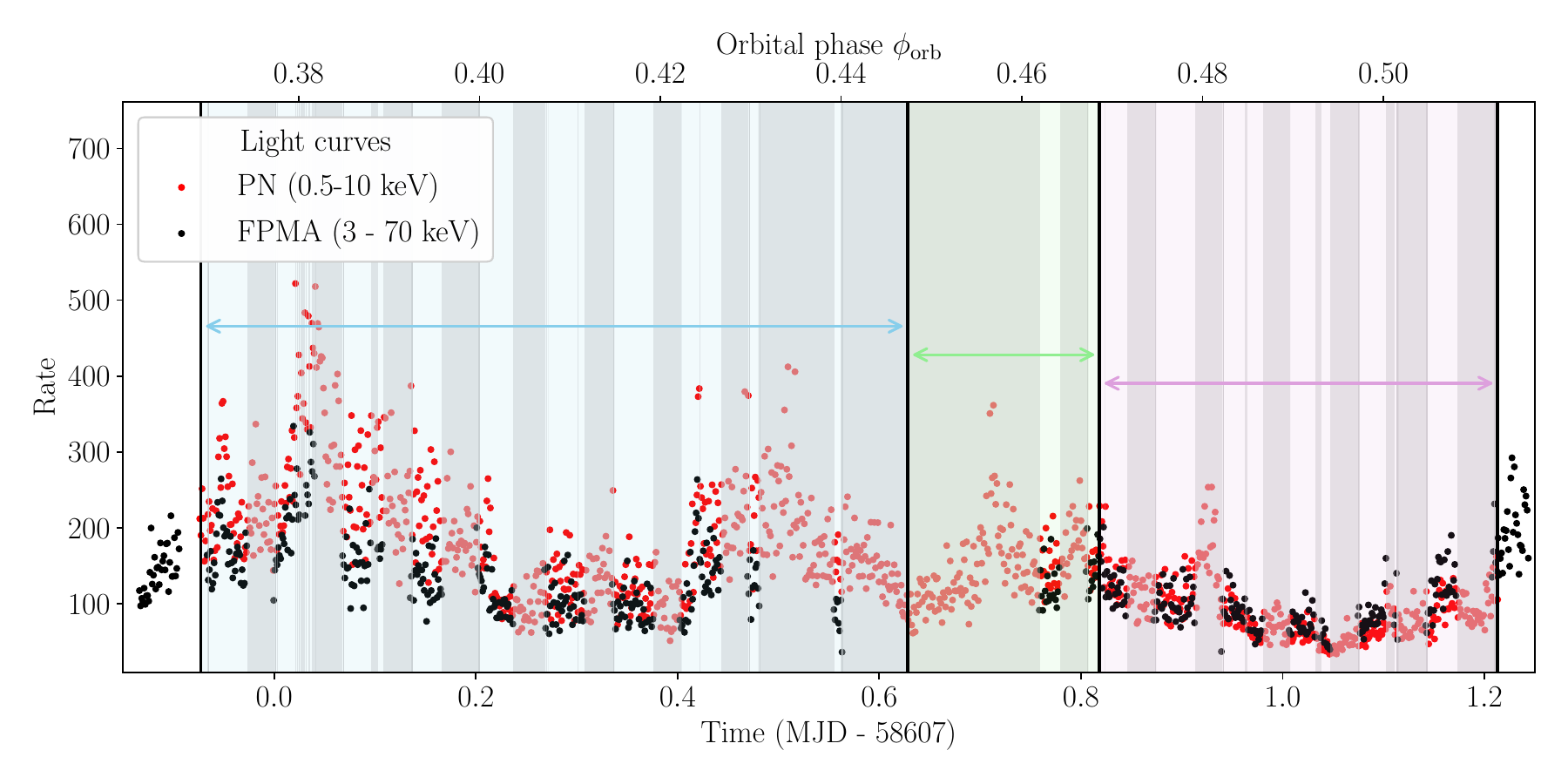}
           \caption{Light curves of \textit{NuSTAR}/FPMA and \textit{XMM-Newton}/EPIC-pn, constructed using time bins of 100\,s. The blue, green, and red boxes (first, second, and third intervals) indicate the three orbital phase intervals as defined by \citet{Diez2023}. The gray shaded regions denote the \textit{NuSTAR} observational gaps.}
    
    \label{fig:gtis}
\end{figure*}

Recently, \citet{Diez2023} analyzed a simultaneous \textit{XMM-Newton} and \textit{NuSTAR} observation of Vela~X-1. \citeauthor{Diez2023} demonstrated that the transition from a soft to a hard state is driven by a rise in absorption column density, consistent with the accretion wake crossing the line of sight. This provided direct evidence for the onset and orbital-phase dependence of the accretion structure in the system. 

In this work we pursue two complementary objectives using the same dataset. First, we introduce the PF spectrum as a quantitative cross-calibration tool by directly comparing energy-resolved pulse profiles extracted from \textit{XMM-Newton}/EPIC-pn and \textit{NuSTAR} over strictly simultaneous time intervals, and we quantify their consistency in the overlapping energy range. Second, we investigate the PF spectrum as a spectro-timing diagnostic probe in its own right by extending its application to the soft X-ray band. Using EPIC-pn data, we derive both the total and orbital phase-resolved PF spectra for the time intervals already defined by \citet{Diez2023} and model their local features within a Bayesian framework guided by the spectral results of that work. In parallel, we analyze the \textit{NuSTAR} data to probe the hard X-ray band and examine potential changes in the pulse profiles at the energies of the CRSFs.


 \section{Observations, data reduction and methodology}


In this section, we build on and update the general framework introduced by \citet{Ferrigno2023}. We investigate the impact of instrumental and observational effects that can influence the reconstructed pulse profiles. These include count-rate-dependent corrections, energy redistribution effects, and the choice of the source and background regions. By quantifying the contribution of all these effects, we aim to assess the residual level of systematics and establish the PF spectrum as a valuable instrumental cross-calibration tool.

\subsection{Observations and data extraction}
\label{subsec:observations}

We used the simultaneous \textit{XMM-Newton} and \textit{NuSTAR} dataset already analysed by \citet{Diez2022} and \citet{Diez2023}. The campaign took place between 3 and 5 May 2019, with \textit{XMM-Newton} ObsID~0841890201 (109.3 ks of net exposure) and \textit{NuSTAR} ObsID~30501003002, 40.6 ks of net exposure. We show the light curves from both observations in Fig.~\ref{fig:gtis}, along with the good-time-intervals (GTIs) of \textit{NuSTAR} and the three orbital phase intervals, as defined by \citet{Diez2023}. 



In the case of \textit{NuSTAR}, we extracted the data using the \texttt{nustardas} package (v1.9.7) within \texttt{HEASoft} (v6.34), together with the calibration files (CALDB v20250224), and produced calibrated level 2 event files for both FPMA and FPMB. We tested different circular source regions around the point-spread function centered on the best-known coordinates of the target, and generated PF spectra for each, in order to quantify the effect of the source region selection of the PF values (see Appendix~\ref{regions_appendix}). The background region was chosen as an equivalent, uniform, area of the detector as far away as possible from the source position. This observation is not affected by stray-light contamination.

For \textit{XMM-Newton}, we reprocessed the data using the Science Analysis System (\texttt{SAS}; v22.1.0) together with the most recent calibration files. We generated calibrated and clean event files using the standard (\texttt{epproc}) command and applied the usual screening on event patterns and quality flags. We used \texttt{evselect} to extract events from a rectangular region centered on the target position along the RAWX direction, while the background was taken from a source-free region on the same CCD. \citet{Diez2023} excluded the four central columns to mitigate pile-up. We explicitly tested the impact of this choice on the PF spectrum by comparing the PF derived from the central columns with that obtained from the remaining source region. We find that the overall shape of the PF spectrum remains essentially unchanged when the central columns are excluded, while the PF amplitude decreases by about $5\%$. Since our goal is to maximize the pulsed signal while preserving the spectral shape, we therefore adopt the full RAWX extraction region for the pulse profile analysis. We refer to Appendix~\ref{regions_appendix} for a detailed discussion of the extraction region selection. Finally, the centroid of the Fe~K$\alpha$ line was measured at $\sim$6.53 \,keV with \textit{XMM-Newton}, significantly higher than the expected rest energy of 6.4 keV. This indicates a residual offset in the EPIC-pn energy scale, as seen in \citet{Diez2023} and \citet{diez2025first} for another \textit{XMM-Newton} observation of the source. To account for this effect, we applied the SAS task \texttt{evenergyshift}\footnote{\url{https://www.cosmos.esa.int/web/xmm-newton/sas-thread-evenergyshift}} to correct the event energies, resulting in a Fe~K$\alpha$ centroid of $\sim$6.44 keV.

\subsection{Pulse profiles extraction}
\label{methodology}

\begin{figure*}
	\centering
	\includegraphics[width=0.49\textwidth]{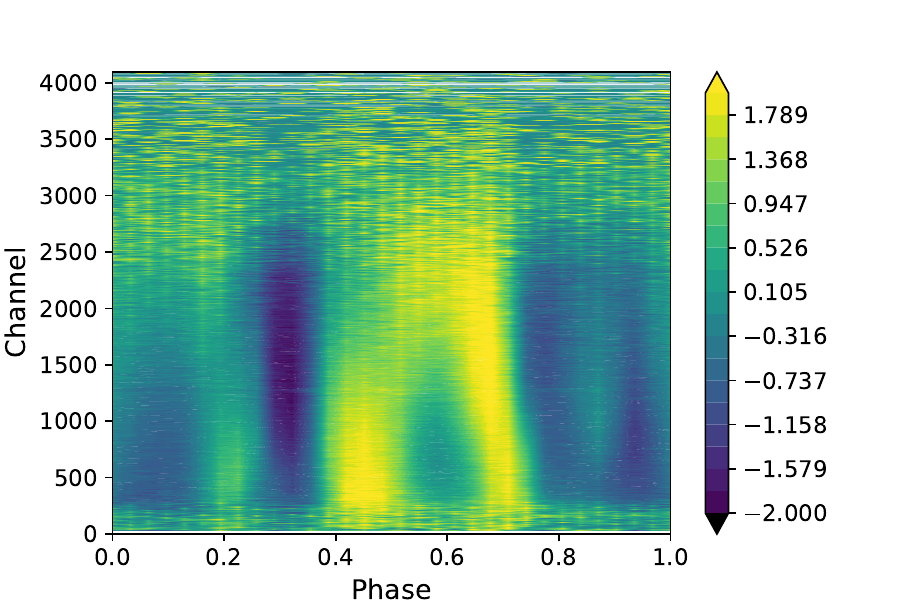}
	\hfill
	\includegraphics[width=0.49\textwidth]{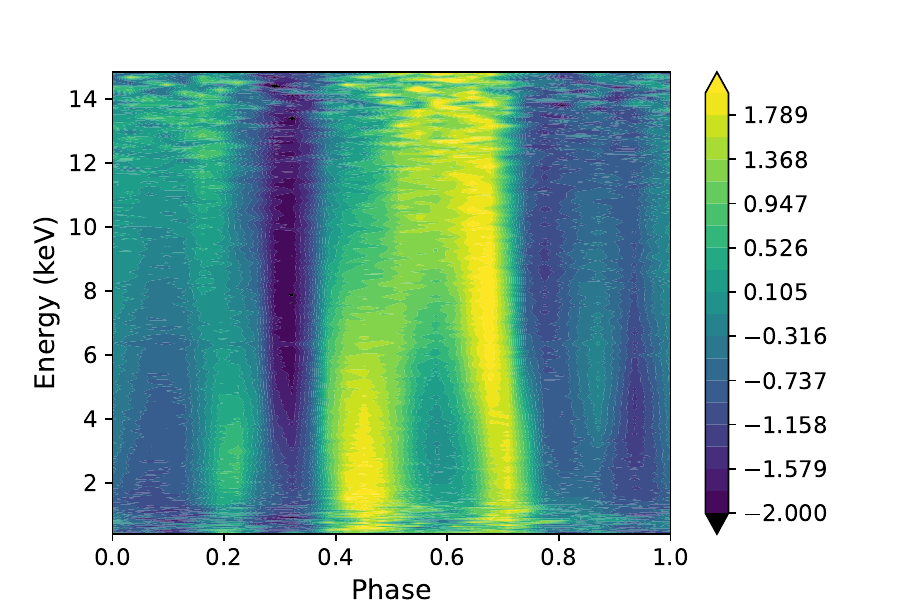}
	\caption{
	(a) Channel--phase matrix of Vela~X-1 obtained with EPIC-pn data using 32 phase bins. 
	(b) Energy--phase matrix derived from the channel--phase matrix after unfolding, rebinning onto the intrinsic EPIC-pn energy grid oversampled by a factor of three, and requiring a minimum signal-to-noise ratio of 15. 
	In both panels, each pulse profile is normalized by subtracting its mean and dividing by its standard deviation.
	}
	\label{fig:matrices}
\end{figure*}

For the timing analysis, we used the reprocessed and cleaned event lists from each instrument. For the comparison presented here, we restricted the analysis to the time intervals common to both datasets and excluded the \textit{NuSTAR} gaps from the EPIC-pn data to ensure identical temporal coverage. After correcting the photon arrival times to the Solar System barycenter, and applying the binary orbital corrections \citep{Malacaria2020}, the events were folded at the best-determined spin period of  $P = 283.4447 \pm 0.0004\,\mathrm{s}$ \citep{Diez2023} at the time of the observation to construct the channel–phase matrices for the EPIC-pn and the \textit{NuSTAR} data, using 32 phase bins.

The exposure of each phase bin is first computed by converting phase intervals into good time intervals. However, the effective exposure can be significantly different because of dead-time corrections,  while the observed count rate is also affected by vignetting and the point-spread function. We apply these corrections and evaluate their impact on the PF spectrum according to the methodology described in Appendix~\ref{exposure_corrections}.

\begin{figure}
    \centering
    \includegraphics[width=0.92
    \columnwidth]{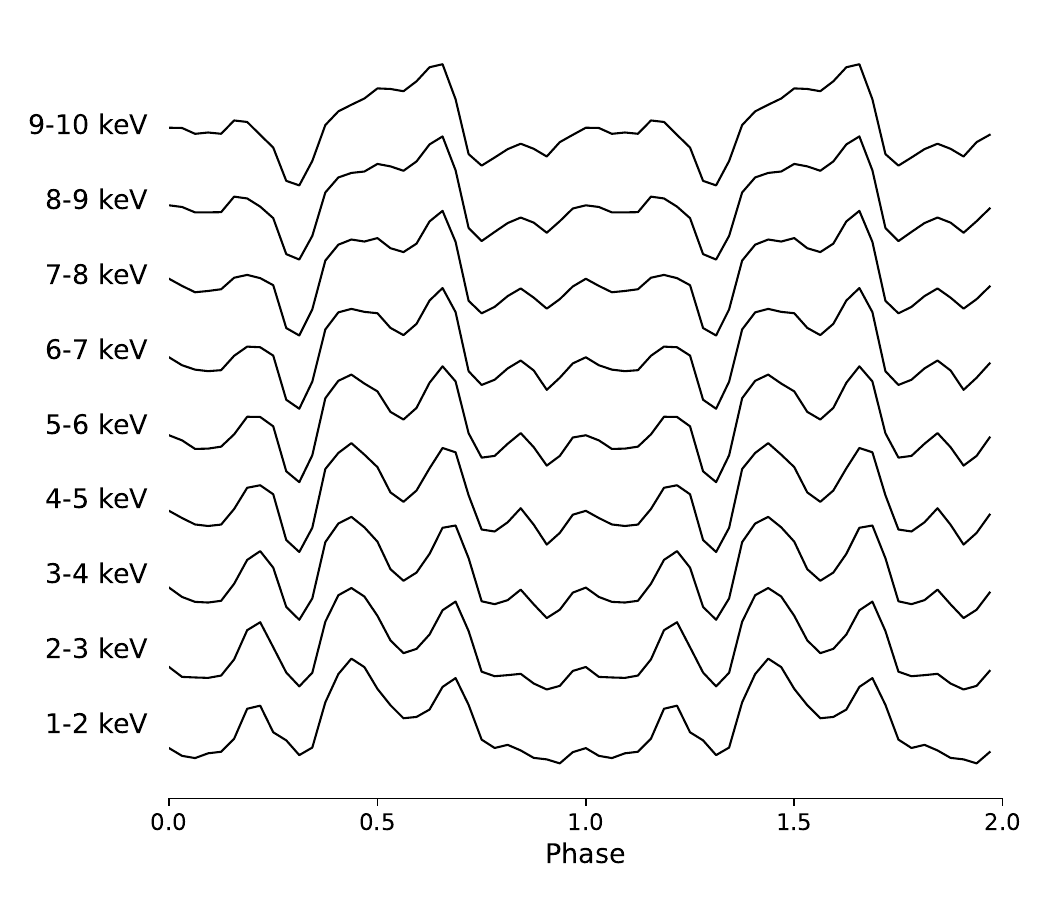}
        \caption{ Energy--resolved pulse profiles with EPIC-pn data, extracted from the energy--phase matrix.
    }
    \label{fig:pps_ref}
\end{figure}

We generated channel–phase matrices independently for the source and the background events. We then subtracted the background channel–phase matrix from the source channel–phase matrix. In the case of \textit{NuSTAR}, the background-subtracted channel–phase matrices from FPMA and FPMB are summed to increase the signal-to-noise (S/N) ratio. The conversion from channel–phase to energy–phase matrices is performed using the \texttt{nDspec} \footnote{\url{https://github.com/nDspec/nDspec}} software package \citep{lucchini2025ndspec}. Following the approach described by \citet{houck2000astronomical}, this procedure allows us to construct unfolded energy–phase matrices expressed in flux density units rather than count rates, thereby consistently accounting for instrumental response effects. A comparison between PF spectra derived from unfolded energy–phase matrices and from count-rate matrices is presented in Appendix~\ref{response_appendix}. Both channel- and energy-phase matrices of EPIC-pn are shown in Fig. \ref{fig:matrices}, while extracted energy--resolved pulse profiles are shown n Fig. \ref{fig:pps_ref}. 

The resulting energy–phase matrices are then re-binned in energy. In the case of EPIC-pn, where sensitivity to narrow spectral features is critical, the energy binning is chosen to match the intrinsic energy resolution of the detector in the 0.5–12\,keV range, oversampled by a factor of three. This choice preserves spectral information without introducing significant biases \citep{Kaastra2016}. For \textit{NuSTAR}, the energy binning is defined following the same principle, oversampling the instrumental energy resolution by a factor of three up to 10\,keV. Above 10\,keV, the binning follows the native energy resolution of the instrument up to 70\,keV. Finally, the matrices are further rebinned to enforce a minimum SNR per energy bin.

PF spectra are finally derived from the re-binned energy–phase matrices using the root-mean-square PF definition,

\begin{equation}
\mathrm{PF}_{\mathrm{rms}} = \frac{1}{\bar{b}}
\sqrt{\sum_{i=1}^{N} \left[ (p_i - \bar{p})^2 - \sigma_{p_i}^2 \right] / N}
,
\end{equation} 

which is equivalent to the Fast Fourier decomposition method \citep{Ferrigno2023}. Here, $N$ is the total number of phase bins, $\bar{p}$ is the average count rate, $p_i$ is the pulse-profile value in the $i$th phase bin, and $\sigma_{p_i}$ is its associated uncertainty. 

Uncertainties on the PF are estimated through a bootstrap procedure. For each energy-resolved pulse profile, we generate 1000 realizations by applying Gaussian fluctuations to the flux in each phase bin. The PF value is computed for every realization, and the standard deviation of the resulting distribution is adopted as the $1\sigma$ uncertainty.

\subsection{Using the PF spectrum as a cross-calibration diagnostic}

\begin{figure}
    \centering
    \includegraphics[width=1.0\columnwidth]{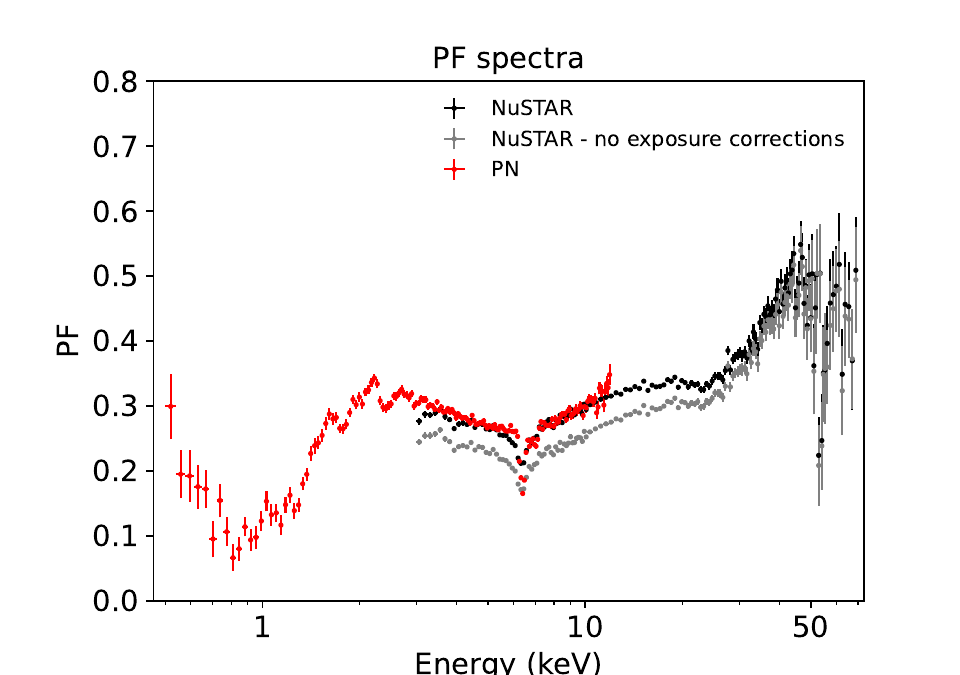}
        \caption{ Pulsed fraction spectra of Vela~X-1 obtained with \textit{NuSTAR} (shown with and without the rate-dependent correction) and with EPIC-pn, using the overlapping exposure time.
    }
    \label{fig:pfs}
\end{figure}

\begin{figure*}
	\centering
	\includegraphics[width=0.49\textwidth]{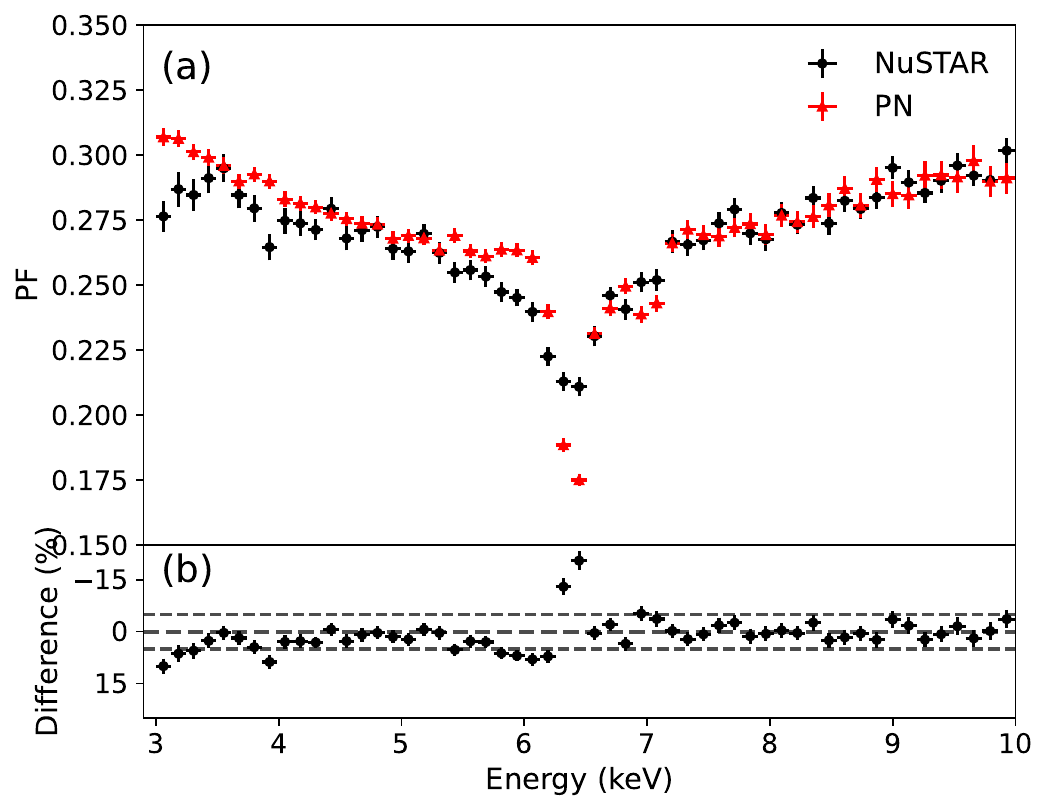}
	\hfill
	\includegraphics[width=0.49\textwidth]{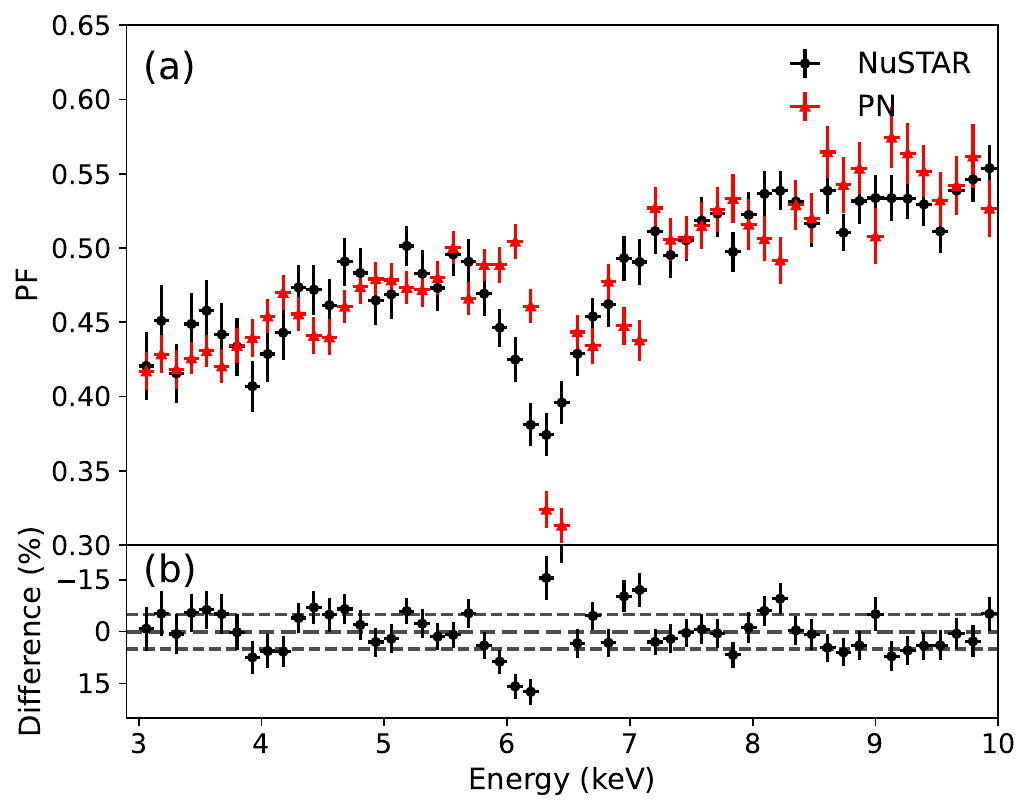}
	\caption{
	Pulsed fraction spectrum calculated using the rms definition (left) and the area definition (right).	Both panels show the PF spectra of Vela~X-1 obtained with \textit{NuSTAR} (black circles) and \textit{XMM-Newton}/EPIC-pn (red triangles) in the overlapping 3--10\,keV range, using a coarser common energy grid. $1\sigma$ uncertainties are displayed. The lower sub-panels in each plot show the percentage difference between the two measurements, computed after interpolating the PF spectra onto a finer common energy grid. Dashed horizontal lines indicate zero difference and the $\pm5\%$ levels.
	}
	\label{fig:pfs_310_comp}
\end{figure*}

\begin{figure}
    \centering
    \includegraphics[width=1.0\columnwidth]{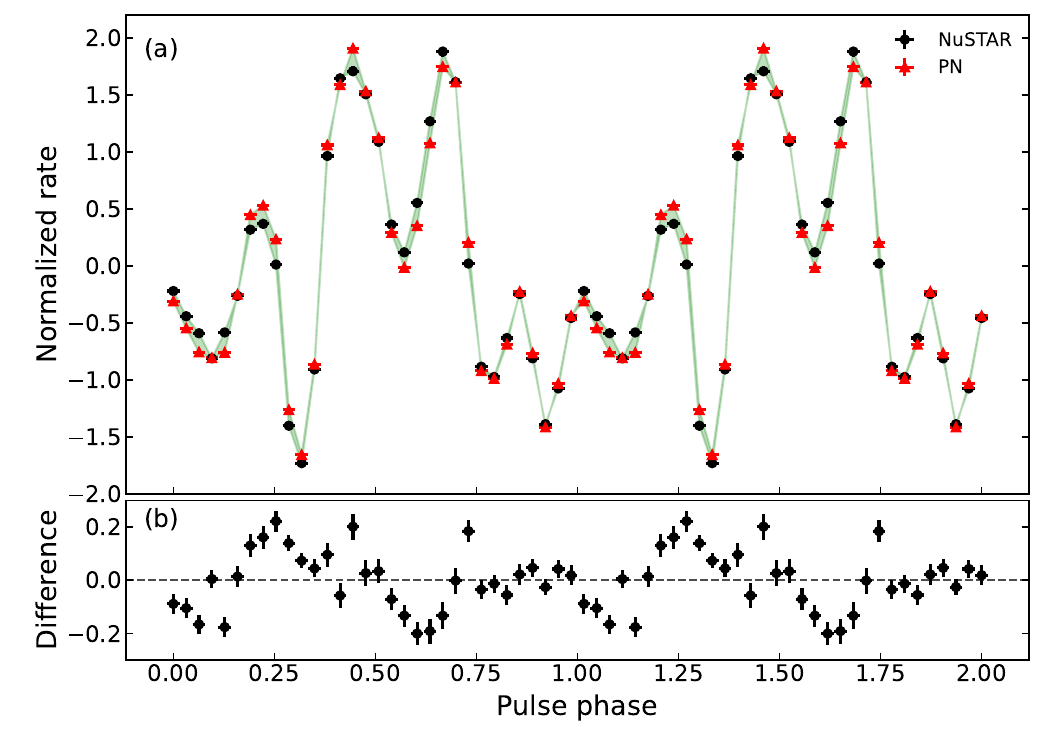}
        \caption{(a) Pulse profiles of Vela~X-1 in the 4--5\,keV energy band extracted with \textit{XMM-Newton}/EPIC-pn and \textit{NuSTAR}, folded into 32 phase bins.  (b) Phase-by-phase difference between the normalized pulse profiles derived from the two instruments. 
    }
    \label{fig:pps36}
\end{figure}

The effect of the source region selection is minimal as long as the majority of the source events are included by the extraction region (see Appendix~\ref{regions_appendix}). A significant difference in the PF values, at the level of $\sim$10\%, is observed only when the central part of the source region is excluded. This test was performed to assess the impact of the piled-up EPIC-pn columns on the PF values \citep{Diez2023}. In this configuration, the PF values decrease for both \textit{NuSTAR} and EPIC-pn. The effect of the background region choice was found to be negligible (less than 1 \%). 

The impact of energy redistribution is discussed in Appendix~\ref{response_appendix}. We found that redistribution effects are more pronounced below 1.5\,keV in the EPIC-pn PF spectra, while remaining negligible in the \textit{NuSTAR} data. The relative difference in these lower energies in the EPIC-pn pulsed fraction reaches at maximum 20\% in just a few energy bins.

The dominant factor affecting the reconstructed pulse profiles of \textit{NuSTAR} is the correction of the count rate for livetime effects \citep{bachetti2026simple}, along with PSF losses and vignetting, as discussed in Appendix~\ref{exposure_corrections}. The PF values obtained without this correction are shown in gray in Fig.~\ref{fig:pfs} and are systematically lower than the corrected ones, with differences reaching up to $\sim$20\%. This demonstrates that livetime effects must be properly accounted for to ensure a reliable cross-calibration of PF spectra. After the correction is applied, the agreement between the \textit{NuSTAR} and EPIC-pn PF spectra improves significantly.

Since the \textit{NuSTAR} energy grid is coarser than that of EPIC-pn, a common energy binning was adopted in order to directly compare the PF values of the two instruments in the overlapping 3--10\,keV band. The \textit{NuSTAR} binning was used as the reference grid, and the EPIC-pn energy--phase matrices were rebinned accordingly by merging adjacent EPIC-pn bins so that the resulting energy intervals match the \textit{NuSTAR} bin boundaries. Since the native EPIC-pn response matrix does not have identical energy boundaries to the adopted grid, small mismatches between the analysis bin edges and the intrinsic EPIC-pn channel boundaries are unavoidable. These differences are typically of order $\sim$10--30 \,eV. 

The PF values measured with the two instruments generally agree within $\sim$5\%. Larger deviations are observed in the iron-line region between 6 and 7\,keV, which can be attributed to the different energy resolutions of the two instruments. Figure~\ref{fig:pfs_310_comp} shows the comparison obtained using both the rms and area definitions of the PF, the latter computed as the phase-integrated excess above the pulse minimum normalized by the total phase-integrated flux, 

\begin{equation}
\mathrm{PF}_{\mathrm{area}} =
\frac{\sum_{i=1}^{N}  \left(p_i - p_{\min}\right)}
     {\sum_{i=1}^{N}  p_i}
\end{equation}

Here, $N$ is the total number of phase bins, $p_i$ is the pulse-profile value in the $i$th phase bin, and $p_{\min}$ is the minimum pulse-profile value among all phase bins. Although the area definition yields systematically larger absolute PF values, the relative agreement between the instruments remains the same.

In Fig.~\ref{fig:pps36}, we show the normalized pulse profiles extracted in the 4--5\,keV energy band using 32 phase bins for both instruments. This narrow energy interval was selected since the effective area of both detectors is relatively flat within this range. By selecting this band, we minimize any instrumental bias on the resulting pulse profile shapes. The lower panel shows the difference between the normalized pulse profiles as a function of phase, illustrating the level of agreement on a phase-bin basis. The deviations remain confined within a narrow range over most of the pulse, indicating good phase-resolved consistency between the two instruments. Despite these local differences, the integrated PF values in the 4--5\,keV band are very similar, with $0.276 \pm 0.001$ for EPIC-pn and $0.271 \pm 0.001$ for \textit{NuSTAR}. The difference corresponds to 2\%, which likely reflects residual cross-calibration systematics between the two instruments and residual imperfect background subtraction. Uncertainties are estimated following the procedure described in Sect.~\ref{methodology}.


\section{Modeling of the Vela~X-1 pulsed fraction spectra}
\label{modelingPFS}

In this section we move from the cross-calibration results to a physical interpretation of the PF spectra. After establishing the robustness of the PF measurements, we investigate their energy dependence to understand how spectral features are reflected in the pulse modulation. The modeling is purely phenomenological and employs a polynomial baseline plus positive and/or negative Gaussian components. The main aim is to characterize the global behavior of the PF spectrum and to identify the presence of local features. Following the methodology introduced by \citet{Ferrigno2023}, we divide the PF spectrum into distinct energy intervals and model each region independently, adopting fitting strategies based on the spectral complexity of each band. We divide the PF spectrum into three energy ranges. The soft band from 1 to 5 keV, covered exclusively by EPIC-pn, contains several X-ray lines, and therefore to avoid continuum–line degeneracies, we adopt a Bayesian framework that allows the use of prior information from spectral studies. The medium band from 5 to 10 keV, where the iron-line complex is found, is covered by both EPIC-pn and \textit{NuSTAR} and is modeled using a least-squares approach on the joint dataset, constructed using only the time intervals strictly simultaneous between EPIC-pn and \textit{NuSTAR}. Finally, the hard X-ray band above 10 keV, covered only by \textit{NuSTAR}, is also modeled using least-squares fitting, following the same approach, with the aim of identifying PF signatures associated with CRSFs.

\subsection{Soft-band modeling (1--5 keV)}
\label{softband}

\begin{figure}
    \centering
    \includegraphics[width=0.9\columnwidth]{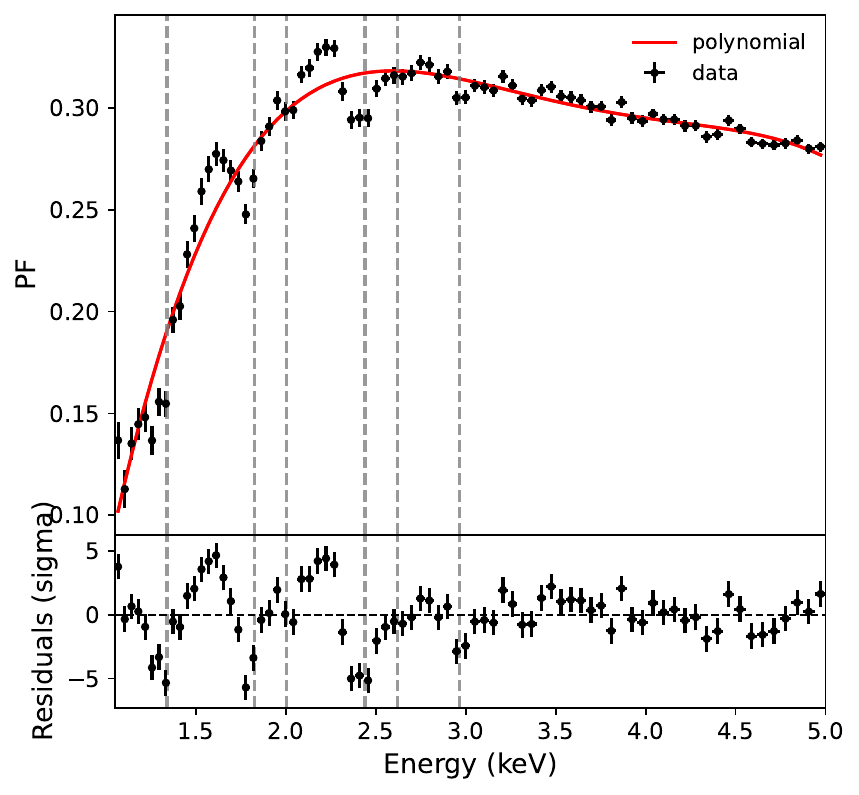}
    \caption{
Polynomial-only fit to the PF spectrum in the 1--5 keV band. The upper panel shows the data together with the best-fit fourth-degree polynomial, while the lower panel shows the residuals. The vertical dashed gray lines indicate the reference line energies from \citet{Diez2023} and are extended to the residual panel to showcase the correspondence between known spectral lines and the depressions observed in the PF.
}
    \label{fig:pfs_onlypolynomial}
\end{figure}

As a first step, we modeled the PF spectrum with a polynomial component to identify possible localized depressions in the PF values that could be associated with spectral features, where clear local dips are observed between $1$ and $3$\,keV (see Fig.~\ref{fig:pfs_onlypolynomial}) 

We assigned Gaussian priors to the centroid energies and line widths: the centroid values were informed by the results of \citet{Diez2023}, while the line widths were constrained using the energy resolution of the EPIC-pn instrument. We performed parameter estimation using MCMC sampling with the No-U-Turn Sampler (NUTS)  implemented in \textsc{PyMC}\footnote{\url{https://www.pymc.io}} \citep{abril2023pymc}. We run four independent chains, each with 4\,000 warm-up (tuning) iterations followed by 4\,000 retained samples, yielding a total of 16\,000 posterior samples per parameter. We assessed the fitting convergence through visual inspection of the trace plots and by requiring the rank-normalized Gelman--Rubin statistic ($\hat{R}$) to be close to unity and the effective sample sizes to be sufficiently large \citep{vehtari2021rank}. Parameter uncertainties were derived from the posterior distributions and are quoted as 1$\sigma$ credible intervals. 

We investigated the presence of localized features in the PF spectrum by comparing models with and without the additional component using Pareto-smoothed importance sampling leave-one-out cross-validation (PSIS-LOO; \citealt{vehtari2017practical}), using the python package \texttt{arviz}\footnote{\url{https://python.arviz.org/}} \citep{kumar2019arviz}.The PSIS-LOO statistic is based on the expected log predictive density (elpd) and provides a measure of the relative predictive performance of competing models while penalizing unnecessary complexity (see, e.g., recent astrophysical applications in \citealt{welbanks2023application, vaughan2024sami, rawlings2025identifying, diez2025first}).

Since the PF spectrum is a derived quantity and neighboring energy bins are intrinsically correlated, PSIS-LOO is used here primarily as a relative model-comparison tool to quantify changes in predictive performance when individual components are removed, rather than as a strict test assuming fully independent data points. The reliability of the PSIS-LOO estimates was assessed through the Pareto-$k$ diagnostic, which measures the influence of individual data points on the importance-sampling approximation \citep{vehtari2017practical}. Values of $k \gtrsim 0.7$ indicate increasingly influential observations, for which the PSIS-LOO approximation may become unreliable, whereas lower values suggest that the cross-validation estimates can be considered robust.


Table~\ref{tab:loo_low_energy} summarizes the best-fit Gaussian parameters and the PSIS-LOO model comparison, reporting the differences in predictive performance, $\Delta \mathrm{elpd}$, with respect to the reference model including all Gaussian components. The $\Delta \mathrm{elpd}$ values quantify the degradation in predictive performance when a given component is removed, while the associated uncertainties (dSE) reflect the statistical uncertainty of the LOO estimate. In practice, components with $\Delta \mathrm{elpd}$ exceeding the associated uncertainty by more than a factor of two are considered to substantially improve the predictive capability of the model, whereas values comparable to or smaller than the uncertainty indicate a weak or negligible contribution. Negative values of $\Delta \mathrm{elpd}$ imply that removing the component improves predictive performance. In the following, we report only components with $\Delta \mathrm{elpd} > 0$, noting that cases with $\Delta \mathrm{elpd} \lesssim \mathrm{dSE}$ are not considered statistically significant.

\subsubsection{Results in the 1 -- 5 keV band}
\label{softresults}

\begin{figure}
    \centering
    \includegraphics[width=0.9\columnwidth]{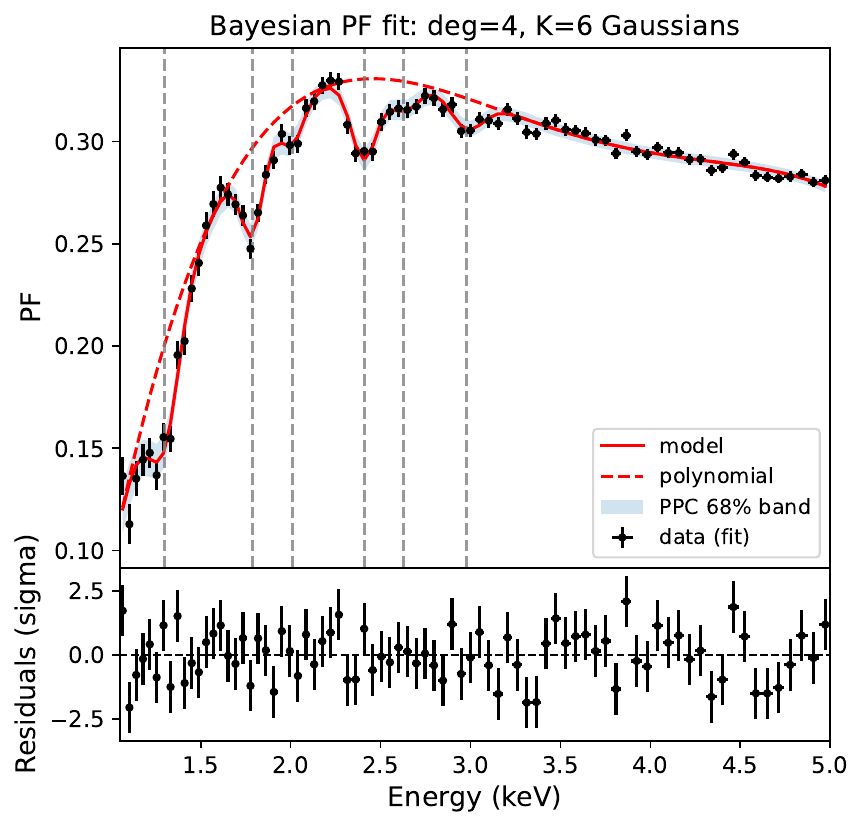}
        \caption{Bayesian fit to the PF spectrum in the 1--5 keV band using a fourth-degree polynomial plus Gaussian components. The upper panel shows the data and the posterior median model, while the lower panel displays the residuals. Shaded regions represent the $1\sigma$ credible intervals of the model. The vertical dashed gray lines mark the reference line energies from \citet{Diez2023}.}
    
    \label{fig:low_band_bayesian_fit}
\end{figure}

We modeled the PF spectrum in the $1$–$5$ keV range using a fourth-degree polynomial to describe the global continuum energy dependence. The data, the best-fit models, and the residuals are shown in Fig. \ref{fig:low_band_bayesian_fit}. We identify 5 significant local dips in the PF spectrum. We associate these features with the blended spectral lines Mg~\textsc{xi} He-like triplet,  Si~\textsc{xiii} He-like triplet, Si~\textsc{xiv} Ly$\alpha$, S~\textsc{xv} triplet and Ar~\textsc{vi}--\textsc{ix} \citep{Diez2023}. After performing the Bayesian fit including these Gaussian components, an additional residual structure appears around $\sim2.6$\,keV, where the S~\textsc{xvi} Ly$\alpha$ line is found in the X-ray spectrum. We note that this feature did not clearly emerge from the residuals of the polynomial-only fit, but only after the other local components are explicitly fitted. We therefore add an additional Gaussian at this energy and performed a final fit of  the data.

We evaluated the significance of each localized feature through the PSIS-LOO model comparison by removing the corresponding Gaussian component from the reference model and quantifying the resulting change in predictive performance (Table~\ref{tab:loo_low_energy}). The removal of the Gaussian components associated with the Mg~\textsc{xi}, Si~\textsc{xiii}, and S~\textsc{xv} He-like triplets results in a large decrease in predictive performance. A similar, though less pronounced, effect is found for the Si~\textsc{xiv} feature, while the Ar~\textsc{vi}--\textsc{ix} and S~\textsc{xvi} Ly$\alpha$ components yields a smaller improvement in predictive capability, suggesting a weaker but still non-negligible contribution. Regarding the reliance of the PSIS-LOO test, although a small number of energy bins exhibit moderately influential behavior ($k>0.7$), no extreme values ($k \ge 1$) are found and the majority of points remain below the conservative threshold of 0.7. Furthermore, the substantial separations in $\Delta \mathrm{elpd}$ between competing models indicate that the ranking is not driven by these moderately influential observations and can therefore be considered robust.

\begin{table}
\centering
\caption{Soft-band (1--5 keV) PF Gaussian components and PSIS-LOO comparison for the total exposure. Uncertainties correspond to $1\sigma$ posterior standard deviations.}
\label{tab:loo_low_energy}
\setlength{\tabcolsep}{3.5pt}
\renewcommand{\arraystretch}{1.15}
\small
\begin{tabular}{lcccc}
\hline
Line & $E_{\rm ref}$ (keV) & $E_{\rm PF}$ (keV) & $\sigma_{\rm PF}$ (keV) & $\Delta\mathrm{elpd}\pm \mathrm{dSE}$ \\
\hline
Mg~XI (f,i,r)
& $1.338^{+0.012}_{-0.019}$
& $1.30 \pm 0.01$
& $0.08 \pm 0.01$
& $50.3 \pm 22.2$ \\

Si~XIII (f,i,r)
& $1.823^{+0.014}_{-0.013}$
& $1.79 \pm 0.01$
& $0.06 \pm 0.01$
& $65.3 \pm 29.4$ \\

Si~XIV Ly$\alpha$
& $2.0049$ (fixed)
& $2.01 \pm 0.02$
& $0.06 \pm 0.02$
& $12.6 \pm 6.7$ \\

S~XV (f,i,r)
& $2.439^{+0.029}_{-0.027}$
& $2.41 \pm 0.01$
& $0.07 \pm 0.01$
& $10.6 \pm 5.0$ \\

S~XVI Ly$\alpha$
& $2.6207$ (fixed) 
& $2.63 \pm 0.03$
& $0.08 \pm 0.02$
& $9.8 \pm 4.9$ \\

Ar~VI--IX
& $2.9661$ (fixed)
& $2.98 \pm 0.02$
& $0.09 \pm 0.02$
& $17.7 \pm 8.9$ \\
\hline
\end{tabular}
\end{table}







\subsection{Medium-band modeling (5--10 keV)}
\label{mediumband}


In this energy range, we fit the joint EPIC-pn/NuSTAR PF spectrum to characterize the global PF behavior across the iron-line band, using time-selected PF spectra from the strictly simultaneous observational intervals of the two instruments. We used the \texttt{lmfit} package to perform least-squares fits to the 5--10 keV PF spectrum, with parameter uncertainties estimated via MCMC sampling using \texttt{emcee} \citep{Foreman-Mackey2013}. Sampling was performed using the affine-invariant ensemble sampler, with 50 walkers, 500 burn-in steps, and 50\,000 steps per walker. We set the MCMC chain length by requiring the total number of steps to exceed the integrated autocorrelation time of the slowest-mixing parameter by at least a factor of 50 \citep{Foreman-Mackey2013}. Best-fit values correspond to the median of the posterior distributions, with uncertainties given by the 16th and 84th percentiles.

\subsubsection{Results in the 5 -- 10 keV band}
\label{mediumresults}

\begin{figure}
    \centering
    \includegraphics[width=0.9\columnwidth]{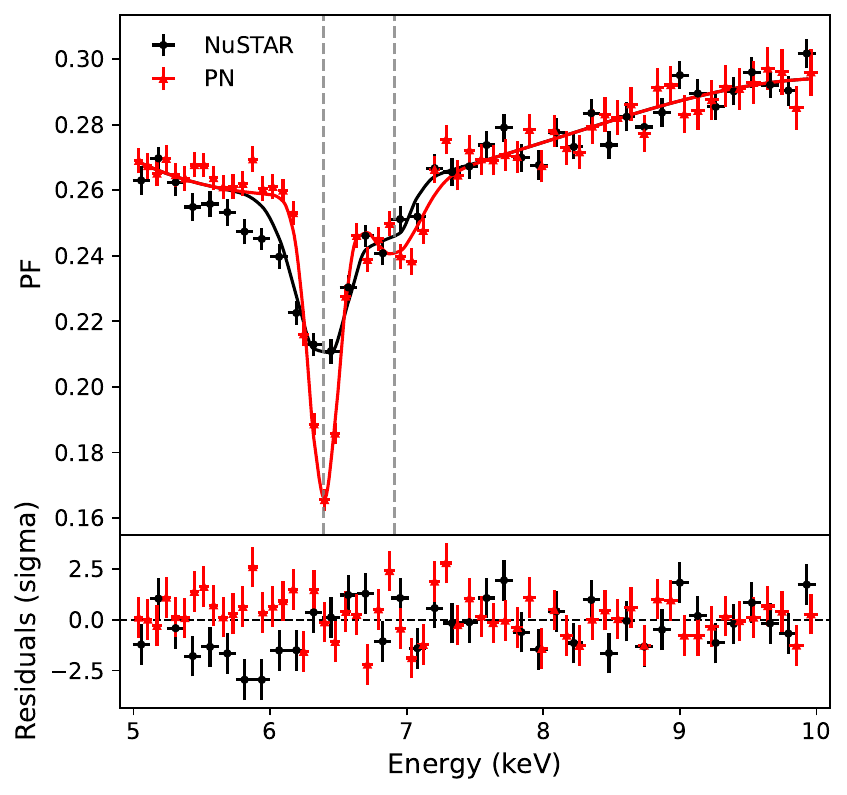}
        \caption{Joint least-squares fit to the 5–10 keV PF spectrum for \textit{NuSTAR} (black) and EPIC-pn (red). The model consists of a third-degree polynomial plus two negative Gaussian components. Residuals are shown in the lower panel in units of $\sigma$. Vertical dashed lines indicate the fitted Gaussian centroids.}
    
    \label{fig:medium_band_joint_fit}
\end{figure}

The PF spectrum in the 5--10 keV band was first modeled using a third-degree polynomial to describe its smooth energy dependence. The degree of the polynomial was determined adaptively within the full model by requiring that the best-fit solution achieves a p-value greater than 5\% in order to balance model accuracy and complexity ($p > 0.05$). The polynomial-only fit provides a very poor description of the data, yielding $\chi^2 = 1608$ for 94 degrees of freedom ($dof$). The inclusion of a negative Gaussian component at 6.4 keV, corresponding to the Fe~\textsc{K}$\alpha$ line, substantially improves the fit, reducing the statistic to $\chi^2 = 224.4$ for 89 dof. Some residual structures remain around 7 keV, and we therefore include a second Gaussian component to account for the possible presence of the Fe~\textsc{K}$\beta$ feature, as reported in the spectral analysis by \citet{Diez2023}. However, the centroid energy of this second Gaussian was not compatible with  the nominal energy of the Fe~\textsc{K}$\beta$ feature (7.06 keV). If kept frozen to this value, the fit yields $\chi^2 = 173.2 $ for 85 dof. If left free, the centroid converges to $\sim 6.9$ keV with a significant fit improvement ($\chi^2 = 128$ for 85 dof). The fit is shown in Fig.~\ref{fig:medium_band_joint_fit}, and the corresponding best-fit parameters are reported in Table~\ref{tab:medium_joint_fit}.

\begin{table}
\centering
\caption{Best-fit Gaussian parameters for the 5--10 keV PF spectrum obtained from the joint least-squares fit. Uncertainties correspond to $1\sigma$ errors from the joint \texttt{emcee} sampling.}
\label{tab:medium_joint_fit}
\setlength{\tabcolsep}{5pt}
\renewcommand{\arraystretch}{1.15}
\small
\begin{tabular}{lccc}
\hline
Component & Parameter & NuSTAR & EPIC-pn \\
\hline
Fe~\textsc{K}$\alpha$ 
& $E$ (keV)     & $6.393 \pm 0.005$ & -- \\
& $\sigma$ (keV)& $0.195^{+0.004}_{-0.008}$ & $0.110 \pm 0.005$ \\
& $A$           & $-0.025 \pm 0.002$ & $-0.026 \pm 0.001$ \\
\hline
Gaussian 2
& $E$ (keV)     & $6.91 \pm 0.03$ & -- \\
& $\sigma$ (keV)& $0.13 \pm 0.04$ & $0.18 \pm 0.02$ \\
& $A$           & $-0.005 \pm 0.002$& $-0.010 \pm 0.002$ \\ \\
\hline
\end{tabular}
\end{table}

Residuals persist around 6 keV in the \textit{NuSTAR} data. When modeling the\textit{NuSTAR} PF spectrum alone, these residuals almost vanish, as the polynomial continuum can better adapt exclusively to the\textit{NuSTAR} data. We therefore attribute the residual structure in the joint fit to a marginal mismatch between the overall PF continua of the two instruments, likely due to their different energy resolutions, which cannot be fully taken into account by allowing only the Gaussian width to vary independently.

In addition, residuals remain around 7 keV in the EPIC-pn data. Even though in \citet{Diez2023}, they do not report an additional feature, observations from other instruments provide evidence for the He-like Fe~\textsc{xxv} complex at $\sim 6.7$ keV \citep{Amato2021, diez2025first}. This suggests that the second Gaussian component may be approximating multiple features rather than representing a single line. To investigate the hypothesis, we model the EPIC-pn spectrum, using the full \textit{XMM-Newton} exposure, within the Bayesian framework introduced in the previous subsection.

\subsubsection{Bayesian re-analysis of the full-exposure EPIC-pn data}
\label{bayesian_medium}

\begin{figure}
    \centering
    \includegraphics[width=0.9\columnwidth]{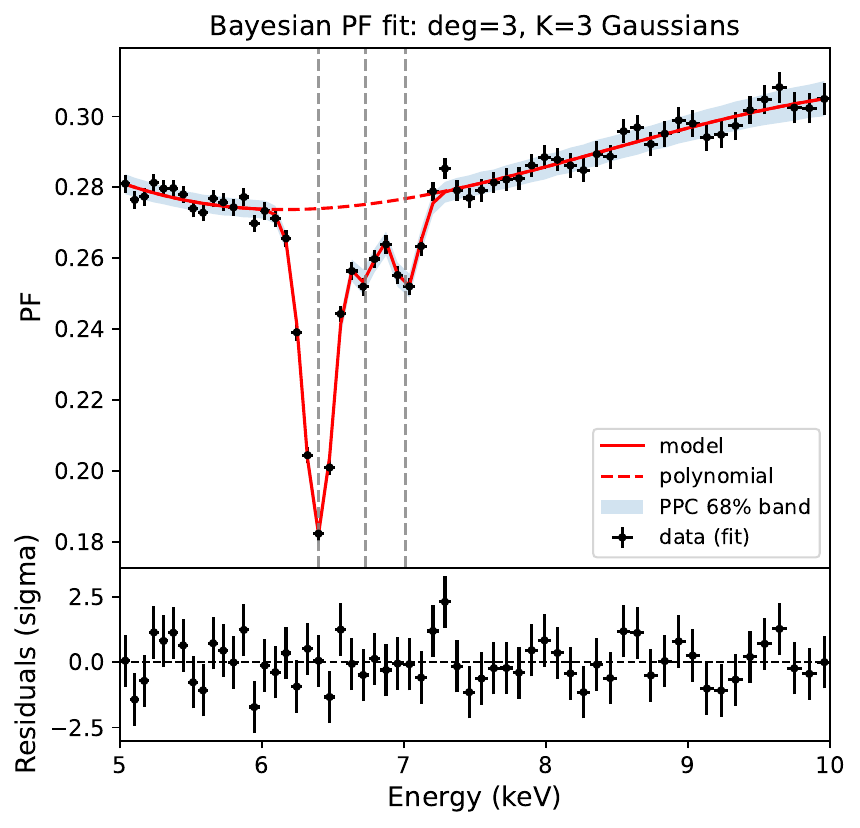}
        \caption{Bayesian modelling of the full-exposure EPIC-pn PF spectrum in the 5--10 keV band using a third-degree polynomial plus Gaussian components. The upper panel shows the data and the posterior median model, while the lower panel displays the residuals. Shaded regions represent the $1\sigma$ credible intervals of the model. The vertical dashed gray lines mark the three iron features.}
    
    \label{fig:medium_band_bayes_fit}
\end{figure}

We used a third-degree polynomial, consistent with the least-squares fit of the joint data, together with three Gaussian components centered at 6.4, 6.7, and 7.06 keV, corresponding to the Fe~\textsc{K}$\alpha$, He-like Fe~\textsc{xxv} complex, and Fe~\textsc{K}$\beta$ spectral features, respectively. When the prior width imposed on the centroid of the 6.7 keV feature was set equal to the instrumental energy resolution, the component converged toward the much stronger 6.4 keV line. To prevent this degeneracy, we imposed a narrower prior on the centroid of the 6.7 keV component, equal to half of the energy resolution, ensuring that its allowed range does not overlap with the 6.4 keV line. The fit is shown in Fig.~\ref{fig:medium_band_bayes_fit}, and the corresponding best-fit parameters are reported in Table~\ref{tab:medium_bayes_fit}.


\begin{table}
\centering
\caption{Fe K-band (5--10 keV) PF Gaussian components and PSIS-LOO comparison for the total exposure of EPIC-pn. Reference energies correspond to the laboratory energies of the respective transitions.}
\label{tab:medium_bayes_fit}
\setlength{\tabcolsep}{3.5pt}
\renewcommand{\arraystretch}{1.15}
\small
\begin{tabular}{lcccc}
\hline
Line & $E_{\rm ref}$ (keV) & $E_{\rm PF}$ (keV) & $\sigma_{\rm PF}$ (keV) & $\Delta\mathrm{elpd}\pm \mathrm{dSE}$ \\
\hline
Fe~K$\alpha$ & 6.400 (fixed) & $6.399 \pm 0.005$ & $0.106 \pm 0.006$ & $1395 \pm 575$ \\
Fe XXV & 6.700 (fixed) & $6.73 \pm 0.03$ & $0.08 \pm{0.03}$ & $61.7 \pm 30.1$ \\
Fe K$\beta$ & 7.058 (fixed) & $7.01 \pm 0.02$ & $0.09 \pm 0.02$ & $76.1 \pm 38.3$ \\
\hline
\end{tabular}
\end{table}

The PSIS-LOO model comparison was performed by sequentially removing the Fe~\textsc{K}$\alpha$, He-like Fe~\textsc{xxv} complex, and Fe~\textsc{K}$\beta$ components from the reference model and evaluating the resulting change in predictive performance (Table~\ref{tab:medium_bayes_fit}). Since the iron features lie close in energy and partially overlap within the instrumental resolution, their parameters are inherently correlated, even with the narrower prior of the 6.7 keV feature. This behavior reflects a clear degeneracy among the components in this spectral region. Therefore, the centroids of the retained Gaussian components were fixed to their best-fit values during the removal tests. Under this controlled configuration, the removal of the Fe~\textsc{K}$\alpha$ line results in a dramatic decrease in $\mathrm{elpd}$. The removal of the He-like Fe~\textsc{xxv} complex and the Fe~\textsc{K}$\beta$ component also reduces predictive performance, though at a more moderate level. While a small number of energy bins exhibit moderately influential behavior ($k>0.7$), most remain below the threshold, no extreme values are present ($k \ge 1$), and the model ranking can therefore be considered robust.

\subsection{Hard-band modeling (10--70 keV)}
\label{modeling_nustar}

\begin{figure}
    \centering
    \includegraphics[width=0.9\columnwidth]{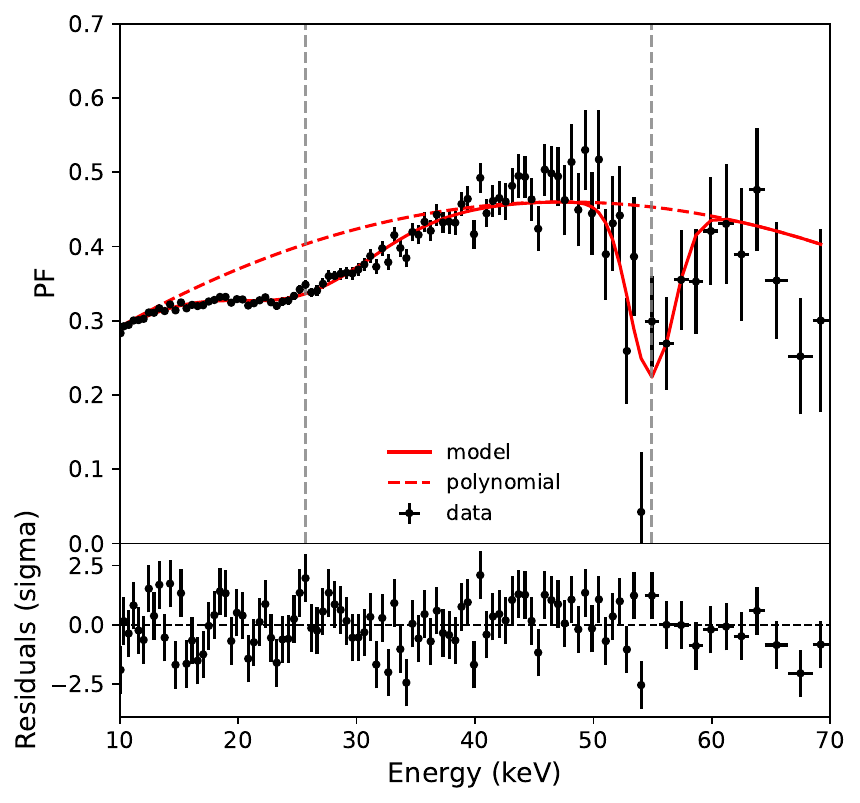}
        \caption{The pulsed fraction spectrum obtained from the full time-averaged \textit{NuSTAR} observation in the 10–70 keV energy range. We describe the continuum with a first-degree polynomial, with two negative Gaussian absorption features at $\sim$\,25 and $\sim$\,55 keV, corresponding to the energies of the cyclotron resonant scattering features identified in spectral analyses.
    }
    \label{fig:hard-band-fit}
\end{figure}

For the hard-band modeling, we use the full-exposure \textit{NuSTAR} data in the 10–70 keV range. The PF spectrum shows a general increase with energy across the full \textit{NuSTAR} band, with the exception of a flattening around 15–25 keV and a pronounced dip at $\sim$\,55 \,keV. These energy ranges coincide with previously reported CRSFs \citep{furst2014nustar, Diez2022, Diez2023}, at approximately 23.5 keV and 55 keV. We therefore added to the fitting model one Gaussian component for each feature. The degree of the polynomial was determined following the least-squares procedure described in the previous subsection, namely through adaptive selection within the full model by requiring that the best-fit solution attains a p-value greater than 5\% ($p > 0.05$). This results in a second-degree polynomial continuum. A fit including only the polynomial provides a poor description of the data, yielding $\chi^2 = 416.3$ for 96 dof. The inclusion of a Gaussian component at $\sim$\,25 keV substantially improves the fit, reducing the $\chi^2 = 138.9$ for 93 dof. Adding the second Gaussian at $\sim$\,55 \,keV further improves the fit, leading to $\chi^2 = 105.1$ for 90 dof. The best-fit model is shown in Fig.~\ref{fig:hard-band-fit}.

The centroid energies of the two Gaussian features inferred from the PF spectrum are fully consistent with the CRSFs reported from phase-averaged spectral analyses of Vela~X–1. The lower-energy feature is centered at $E_{1}=25.6_{-0.2}^{+0.2}$\,keV with a width $\sigma_{1}=5.9_{-0.2}^{+0.1}$\,keV, consistent with the fundamental CRSF observed at $\sim$24–25\,keV in time-averaged \textit{NuSTAR} spectra for the same observation \citep{Diez2022}. While previous spectral analyses reported that the width of the fundamental CRSF could not be robustly constrained \citep{Fuerst2014,Diez2022,Diez2023}, both the centroid energy and width are well determined in the PF spectrum. The higher-energy feature is centered at $E_{2}=55.1_{-0.5}^{+0.6}$\,keV with a width $\sigma_{2}=2.1_{-0.5}^{+0.6}$\,keV, consistent with the harmonic CRSF whose centroid varies between $\sim$\,52 and $\sim$\,56\,keV across different \textit{NuSTAR} observations  \citep[e.g.,][]{Diez2022}.


\section{Orbital-phase-resolved modeling of the soft-band PF spectrum with EPIC-pn}

We also fit the three orbital-phase-resolved PF spectra, using the time intervals defined by \citet{Diez2023}: during pre-wake, accretion wake ingress and wake-dominated phase (see Fig.~\ref{fig:gtis}), with the aim of studying the behavior of the PF values in the soft-band, where absorption due to the accretion wake has an overall significant effect also for the pulse shapes. The corresponding best-fit parameters are reported in Table~\ref{tab:loo_low_energy_orbital}.

\begin{table*}
\centering
\caption{Soft-band (1--5 keV) PF Gaussian components and PSIS-LOO comparison for the pre-wake and accretion wake ingress phases. Uncertainties correspond to $1\sigma$ posterior standard deviations.}
\label{tab:loo_low_energy_orbital}
\setlength{\tabcolsep}{4pt}
\renewcommand{\arraystretch}{1.15}
\small
\begin{tabular}{lccccccc}
\hline
Line 
& $E_{\rm ref}$ 
& \multicolumn{3}{c}{eclipse egress} 
& \multicolumn{3}{c}{inferior conjunction} \\
\cline{3-5} \cline{6-8}
& (keV)
& $E_{\rm PF}$ 
& $\sigma_{\rm PF}$ 
& $\Delta\mathrm{elpd}\pm\mathrm{dSE}$
& $E_{\rm PF}$ 
& $\sigma_{\rm PF}$ 
& $\Delta\mathrm{elpd}\pm\mathrm{dSE}$ \\
\hline

Mg~XI (f,i,r)
& $1.338^{+0.012}_{-0.019}$
& $1.28 \pm 0.01$
& $0.06 \pm 0.01$
& $16.5 \pm 7.0$
& $1.41 \pm 0.02$
& $0.06 \pm 0.02$
& $7.2 \pm 5.0$ \\

Si~XIII (f,i,r)
&  $1.823^{+0.014}_{-0.013}$
& $1.79 \pm 0.01$
& $0.04 \pm 0.01$
& $21.6 \pm 17.0$
& $1.77 \pm 0.03$
& $0.10 \pm 0.02$
& $15.1 \pm 7.2$ \\

Si~XIV Ly$\alpha$
& $2.0049$ (fixed)
& $2.02 \pm 0.04$
& $0.06 \pm 0.02$
& $1.2 \pm 1.9$
& -- 
& -- 
& -- \\

S~XV (f,i,r)
& $2.439^{+0.029}_{-0.027}$
& $2.43 \pm 0.01$
& $0.08 \pm 0.01$
& $3.7 \pm 2.8$
& $2.41 \pm 0.03$
& $0.06 \pm 0.02$
& $2.7 \pm 1.9$  \\

Ar~VI--IX
& $2.9661$ (fixed)
& $2.98 \pm 0.03$
& $0.07 \pm 0.02$
& $27.8 \pm 12.6$
& -- 
& -- 
& -- \\

\hline
\end{tabular}
\end{table*}








In the low-energy band, the PF spectrum extracted from the first interval exhibits a global behavior similar to that of the time-averaged PF spectrum, although with systematically higher PF values across the band. In this orbital phase, clear localized dips are present at energies consistent with the features identified in the time-averaged analysis, and we model them using Gaussian components. The PSIS-LOO model comparison shows that the Gaussian components associated with the Mg~\textsc{xi} and Ar~\textsc{vi}--\textsc{ix} improve the predictive performance of the model. The Gaussian components associated with the Si~\textsc{xiii}, Si~\textsc{xiv}, and S~\textsc{xv}  provide a more moderate improvement. In contrast, adding a Gaussian component at $\sim2.6$ keV, corresponding to S~\textsc{xvi} Ly$\alpha$, worsen the predictive performance and therefore we do not include it in the final model. 

In the PF spectrum extracted from the second interval, structures at $\sim1.3$, $\sim1.8$, and $\sim2.4$\,keV are visible in the PF spectrum. Even though the Gaussian component associated with the S~\textsc{xv} feature at $\sim2.4$\,keV remains consistent with the expected transition energy, the components associated with Mg~\textsc{xi} ($\sim1.3$\,keV) and Si~\textsc{xiii} ($\sim1.8$\,keV) show centroid energies that are more displaced from the corresponding reference transitions compared to the first PF spectrum, particularly in the case of Mg~\textsc{xi} whose centroid was already affected by the continuum shape. This behavior likely reflects the drop of the PF values as a consequence of the increased column density, which results in a noisier spectrum. The Gaussian components are probably modeling statistical fluctuations rather than tracing well-defined localized dips. 

The situation is even more severe in the third, high-absorption interval, where the PF values remain extremely low up to $\sim2.5$\,keV, preventing a meaningful modeling of the $1$--$5$\,keV band, and therefore we do not attempt to model this PF spectrum.

\begin{figure}
    \centering
    \includegraphics[width=0.9\columnwidth]{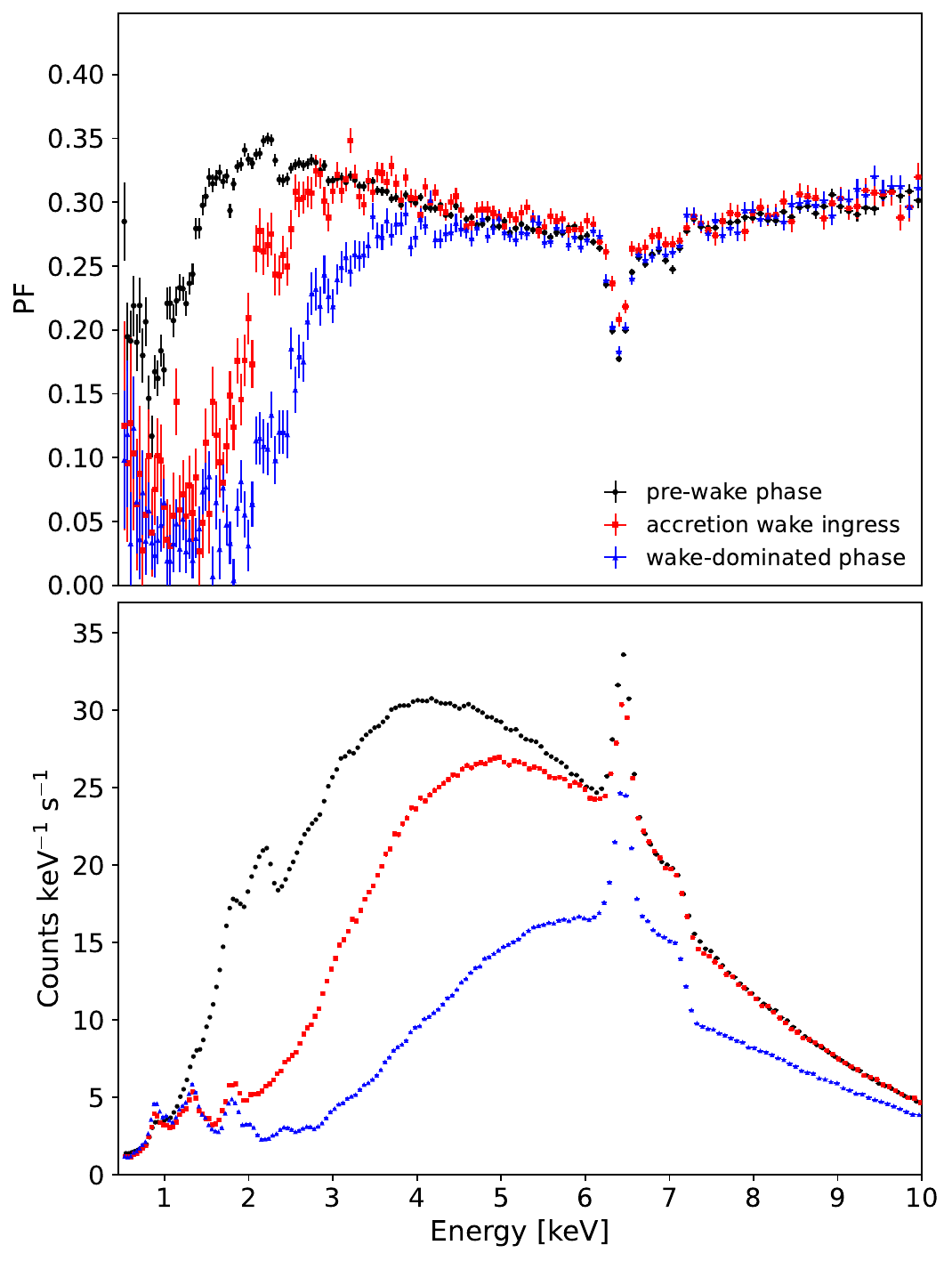}
        \caption{
        Pulsed fraction spectra (top panel) and corresponding orbital-phase-resolved X-ray energy spectra (bottom panel) in the soft-band for the three GTIs: pre-wake, accretion wake ingress and wake-dominated phase.} 
        
    \label{fig:all3}
\end{figure}

In Fig.~\ref{fig:all3} we show the three orbital-phase-resolved energy spectra together with the corresponding PF spectra. In the orbital-phase-resolved X-ray spectra, the emission lines are more pronounced during the third orbital interval, which is characterized by a more absorbed continuum. In contrast, the opposite behavior is observed in the PF spectra, where the lines, detected as local dips, appear strongly suppressed in the third interval and much more pronounced during the first interval.

\section{Discussion}

\subsection{Cross-calibration via energy-resolved pulse profiles}

A direct comparison of pulse profiles obtained with different instruments requires careful control of instrumental and observational effects that can modify the reconstructed pulse shape. By constructing PF spectra from simultaneous \textit{XMM-Newton}/EPIC-pn and \textit{NuSTAR} observations of Vela~X-1, we quantified the impact of these effects and assessed the consistency of the energy-resolved pulse profiles measured by the two missions. 

The most significant effect in the PF values of the \textit{NuSTAR} data is related to the correction of the phase-dependent count-rate. To account for this, we applied the correction described in Appendix~\ref{exposure_corrections}, which adjusts the effective exposure of each phase bin. This correction is largely driven by livetime effects, which are particularly important for bright sources such as Vela~X-1, but it also incorporates the effects of the point-spread function and vignetting. Without applying this correction, the PF derived from the \textit{NuSTAR} data is systematically underestimated by up to $\sim20\%$ in the overlapping energy range. After applying this correction, the PF spectra of \textit{NuSTAR} and EPIC-pn become consistent within a few percent in the 3--10\,keV band. 

We also examined the impact of the source extraction regions. In particular, we tested the effect of excluding the central EPIC-pn columns that may be affected by pile-up. While pile-up can modify the reconstructed pulse profile, removing these columns also excludes a large fraction of the source counts and leads to a systematic reduction of the measured PF. Since our goal is a fine comparison of PF spectra between instruments, excluding pile-up columns would introduce a systematic far larger than the marginal pulse shape distortions if the pile-up columns would have been removed. Excluding the central columns would also require adopting a significantly different source selection relative to the \textit{NuSTAR} extraction region, making the comparison between the two instruments less straightforward. For this reason we retain the full EPIC-pn extraction region. As shown in Appendix~\ref{regions_appendix}, the PF spectra obtained from the central columns and from the outer source region show consistent energy dependence.

Other tested effects produce significantly smaller changes. The choice of the background region modifies the PF values by less than $\sim1\%$, while energy redistribution effects remain negligible for \textit{NuSTAR}, however it introduces some differences in the EPIC-pn PF spectrum below $\sim1.5$\,keV. After accounting for these effects, the remaining discrepancies between the two instruments are limited to the iron-line region around 6--7\,keV. These differences are most likely related to the different energy resolutions of the EPIC-pn and \textit{NuSTAR} detectors, which affect how narrow spectral features are redistributed in the energy–phase matrices.

In the case of Vela~X-1, there is a clear orbital-phase dependence of the PF spectrum, indicating the importance of strictly simultaneous observations when using PF spectra as a cross-calibration tool. In general, however, we argue that this method could be reasonably well applied also to non strictly simultaneous observations for more stable sources, provided that the source is in the same spectral shape and the pulse shapes are consistent.

We also argue that this method can provide a complementary route to cross-calibration at hard X-ray energies. In this regime, narrow spectral features suitable for calibration are often weak or absent, so cross-calibration commonly relies on the comparison of continuum shape and normalization, which depend directly on the energy-dependent effective area. In contrast, PF values are relative quantities measured within individual energy bins and, to first order, do not depend on the absolute effective area. PF spectra can therefore help disentangle instrument-dependent effects on the pulse profiles  from uncertainties in the absolute effective area.

\subsection{Timing signatures of soft spectral features}

We used the spectral information obtained from \citet{Diez2023} to guide a Bayesian modeling approach of the PF spectrum. Localized dips are observed at energies corresponding to several emission lines, including soft X-ray lines and iron transitions. The reduced PF at these energies implies that the line photons are not pulsed with respect to the continuum emission, consistent with an origin in reprocessing material located away from the immediate vicinity of the NS. We also modeled orbital-phase–resolved PF spectra in the soft band to investigate how the PF spectrum and the associated line-related dips evolve as the absorption column density increases along the orbit.

According to \citet{Diez2023}, the line-emitting material is distributed throughout the circumstellar environment, while the absorber is localized. The strong absorption observed during the observation is attributed to the accretion wake progressively crossing the line of sight. This interpretation is further supported by the detection of similar soft emission lines during eclipse observations of Vela~X-1 \citep{Sako1999}, when the neutron star and its local absorber are outside the line of sight. The results of the orbital-phase–resolved PF analysis support this scenario. The PF dips are strongest in the first orbital-phase–resolved PF spectrum, where the absorption column density is lowest and the accretion wake has not yet entered the line of sight. In intervals with higher absorption, the PF is strongly suppressed, reducing the detectability of localized PF structures associated with spectral features. This behavior is consistent with an origin of the line-emitting material in the extended stellar wind rather than within the accretion wake, as PF dips associated with these lines are clearly detected before the neutron star becomes embedded in the dense accretion wake.

In addition, we find evidence for a feature at $\sim6.7$\,keV consistent with the He-like Fe~\textsc{xxv} line, which is not detected in the phase-averaged EPIC-pn X-ray spectrum. Its presence in the PF spectrum suggests that some spectral components are strongly phase dependent and can be diluted in phase-averaged spectra. PF analysis can therefore provide complementary constraints to conventional spectral studies.

\subsection{Cyclotron resonant scattering features in the pulsed fraction spectrum}

The timing signatures of CRSFs in PF spectra vary between sources. \citet{Ferrigno2023} reported broad PF dips at the CRSF energies in several accreting pulsars, while \citet{D'Ai2025} found two narrow wing-like bumps in the PF spectrum of V~0332+53 and \citet{maniadakis2025} reported a single broad bump in 4U~1538–52. These differences were interpreted as arising from the interplay between the spin-dependent flux used to compute the PF and the spin-phase variability of the CRSF profile \citep{maniadakis2025}. More recently, \citet{Ambrosi2026} showed that the PF signature of the cyclotron line depends on luminosity in 4U~1901+03, and that the CRSF does not always produce a detectable feature in the PF spectrum.

In our \textit{NuSTAR} analysis of Vela~X-1 we detect PF variations at energies corresponding to both the fundamental CRSF and its harmonic. Interpreting these dips in terms of spin-phase spectral variability suggests that the cyclotron lines may become deeper during phases of higher flux. Indeed, \citet{maitra2013pulse} showed by analyzing \textit{Suzaku} data, that the depth of both CRSFs in Vela~X-1 varies strongly with pulse phase, being largest near the pulse peaks and much shallower in off-pulse phases, consistent with what we would expect from the PF spectrum. Nevertheless, a dedicated spin-phase–resolved analysis would be required to fully interpret the PF spectrum of this dataset with \textit{NuSTAR}, which is beyond the scope of the present work.

However, we note that the Gaussian component used to model the fundamental cyclotron line (at $\approx 25$ \,keV) is very broad and shallow, which seems to indicate a change in the continuum rather than a genuine timing signature of a cyclotron feature. Using an alternative model consisting of a 5th degree polynomial and a single Gaussian component at $\sim55$\,keV yields a fit with $\chi^2 = 95.5$ for 90 dof.
Indeed, the interpretation of the shallow absorption feature as a fundamental cyclotron line in Vela~X-1, as well as in two other sources -- 4U~1907+09 and GX~301$-$2 -- has been questioned previously, potentially being an artifact of the continuum modeling (e.g., \citealt{orlandini2001hard}). For GX~301$-$2, an alternative spectral description was proposed that provides a good fit to the data using a power-law continuum and a pronounced cyclotron line at the energy reported for the second harmonic, together with a broad red wing modeled by a Gaussian emission component \citep{Zalot2024}. In this scenario, the spectrum arises from a formation process similar, though less extreme, to that producing the double-hump spectra observed in low-luminosity states (e.g., \citealt{Tsygankov2019b}, \citealt{Sokolova-Lapa2021}), where the high-energy hump can be attributed to the emission wing of the cyclotron line. PF spectra for a source in such an accretion state have recently been reported by \citet{Malacaria2026}, although the analysis was limited to the energy range between the two humps due to the low signal-to-noise ratio. We therefore stress that further in-depth phase-resolved analysis, supported by physical modeling, is required to clarify the nature of the 25-keV absorption feature in the energy and PF spectra of Vela~X-1.

\section{Conclusions}

In this work, we have performed the first direct and quantitative comparison of energy-resolved pulse profiles obtained with \textit{XMM--Newton}/EPIC-pn and \textit{NuSTAR}, taking advantage of a strictly simultaneous observation of the accreting X--ray pulsar Vela~X-1. By using the PF values as a measure of the pulsation strength in each energy bin, we assessed how instrumental effects modify the pulse profile shapes. Our methodology followed a systematic approach to eliminate instrumental bias by first generating raw channel-phase matrices for both instruments. We then applied corrections for count-rate–dependent effects, primarily driven by livetime, as well as by PSF and vignetting. These corrected matrices were then unfolded from detector channels to physical energy units, a step that allowed us to derive the PF spectra in flux space rather than count space. Finally, we quantified the impact of the source region selection by testing multiple extraction regions for both EPIC-pn and \textit{NuSTAR}. The resulting PF spectra agree at the level of a few percent in their overlapping 3--10 keV energy range. The comparison of PF spectra allowed us to isolate which instrument-specific corrections were responsible for the remaining differences between EPIC-pn and \textit{NuSTAR}. In particular, the count-rate-dependent correction was essential to reconcile the PF values measured by EPIC-pn and \textit{NuSTAR}. This correction had a large effect on the \textit{NuSTAR} PF spectrum, while its impact on EPIC-pn was negligible, demonstrating that different instrumental characteristics can affect the inferred pulse profile properties in different ways. We therefore propose the use of PF spectra as a robust and sensitive observable for comparing energy-resolved pulse profiles across different instruments, as well as a useful diagnostic for cross-calibration.

Extending the PF spectrum analysis over the $1$--$70$\,keV band, we confirm that there are  imprints of local spectral features in the PF spectrum. In the soft X-ray band covered by EPIC--pn, multiple localized dips in the PF spectrum coincide with known emission lines, revealing a complementary timing signature of line-forming regions that has not been explored before. The orbital-phase-resolved PF spectra further show that the detectability of these features depends on the absorption state of the system. Emission-line signatures in the PF spectrum become progressively weaker during more absorbed intervals, indicating that the PF encodes information not only on the emission geometry but also on the surrounding environment. This behavior highlights the potential of PF-based diagnostics to disentangle geometric and environmental effects that are often difficult to separate using spectroscopy alone. In the \textit{NuSTAR} band, PF variations are observed at energies consistent with the fundamental cyclotron resonant scattering feature and its harmonic. While the harmonic appears as a clear narrow dip near $\sim55$\,keV, the structure near the fundamental energy is broad and shallow, and may instead reflect continuum variations rather than a genuine timing signature of a cyclotron line. 

Overall, our results establish the pulsed fraction spectrum as both a powerful spectro-timing diagnostic and a practical cross-calibration tool for X-ray pulsar studies. The methodology developed here can be readily extended to other missions with high timing capabilities, such as \textit{NICER} and \textit{XRISM}, paving the way for unified, broadband pulse-profile studies across the soft and hard X-ray domains.

\section*{Data availability}

All data and software used in this work are publicly available from the ESA and NASA Data Science Archives (\url{https://www.cosmos.esa.int/web/xmm-newton/xsa}), (\url{https://heasarc.gsfc.nasa.gov/}).

\begin{acknowledgements}
The authors are grateful to Antonio Tutone and Jakob Stierhof for their valuable comments and helpful discussions. 
This research was carried out within the European Space Agency (ESA) \href{https://www.cosmos.esa.int/web/esdc/visitor-programme}{Archival Research Visitor Programme}, through which D.K.M. was supported.

A.D. and C.P. acknowledge support from PRIN MUR 2022 SEAWIND, funded by the European Union -- Next Generation EU, Mission 4 Component 1, CUP C53D23001330006, and from the INAF Large Grant 2023 BLOSSOM, F.O. 1.05.23.01.13.

C.M.D. acknowledges support through the ESA Research Fellowship Programme in Space Science.

E.A. acknowledges support from the INAF MINI-GRANTS, grant number 1.05.23.04.04, for the project PPANDA (Pulse Profiles of Accreting Neutron Stars Deeply Analyzed).

E.A., A.D., and G.C. acknowledge funding from the Italian Space Agency under contract ASI/INAF n.~I/004/11/4.

C.M. is supported by INAF (Research Large Grant FANS and GO Grant PULSE-X, PI: Papitto), the Italian Ministry of University and Research (PRIN MUR 2020, Grant 2020BRP57Z, GEMS, PI: Astone), and Fondazione Cariplo/Cassa Depositi e Prestiti (Grant 2023-2560, PI: Papitto).
\end{acknowledgements}

\bibliographystyle{aa}
\bibliography{refs}


\appendix
\titlespacing*{\section}{1pt}{*1}{*1}

\section{Impact of instrumental effects on pulse profile extraction}

\subsection{Rate-dependent corrections}
\label{exposure_corrections}

Different instrumental characteristics and dead-time effects result in a rate-dependent effective exposure, as the live-time depends on the instantaneous count rate over the full detector area. As a consequence, pulse profiles constructed directly from the event files need a phase-dependent exposure correction, whose absence might lead to a systematic underestimate of the pulsed fraction.

\begin{figure}
	\centering
	\includegraphics[width=0.9\columnwidth]{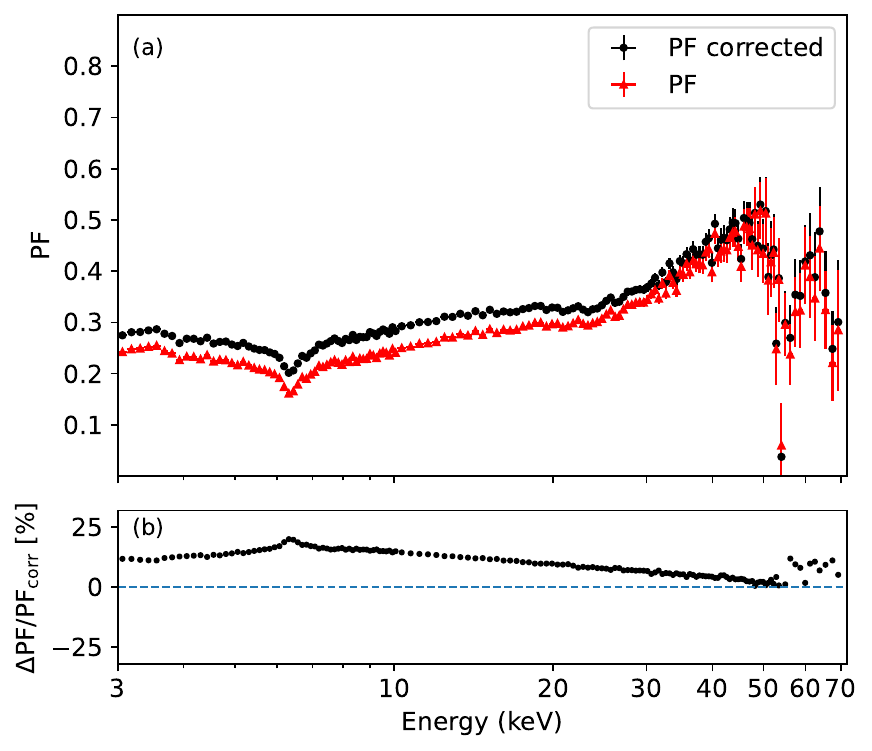}
	\caption{
	Panel~(a) compares the PF spectrum computed with and without applying the rate-dependent correction. Panel~(b) shows the corresponding relative difference between the two cases, expressed as a percentage, over the considered energy range.
}

	\label{fig:exposure_nustar}
\end{figure}

To quantify this effect, we compare the observed count rate $R$ with the exposure-corrected rate $R_{\mathrm{corr}}$, both derived from light curves. In \textit{NuSTAR} \citep{bachetti2015no}, the two quantities show a clear non-linear correlation, reflecting the rate-dependence of the effective exposure. We modeled this relation by fitting $R_{\mathrm{corr}}$ as a function of $R$ with a quadratic function, and used the resulting best-fitting parameters to correct the exposures in each phase bin of the pulse profile. The corrected pulse profiles are then used to compute the PF spectrum. The impact of this correction is illustrated in Fig. \ref{fig:exposure_nustar}. When the $R$--$R_{\mathrm{corr}}$ correction is not applied, the pulsed fraction is systematically underestimated, with discrepancies reaching $\sim 20$--$30\%$ over a broad energy range. 

For \textit{XMM-Newton}/EPIC-pn Timing Mode data, we applied the same procedure described above, deriving the exposure-corrected rate $R_{\mathrm{corr}}$ from the corresponding light curves and fitting it as a function of the observed count rate $R$. In this case, the $R$--$R_{\mathrm{corr}}$ relation is well described by a linear function. As a consequence, applying the correction to the EPIC-pn pulse profiles results to an up-scaling rather than a reshaping, which has negligible effects on the PF values. Nevertheless, we treated the data in a homogeneous way by applying the same correction scheme, ensuring that any residual differences in the PF spectra are not introduced by methodological inconsistencies, but reflect genuine instrumental, or physical, effects.

\subsection{Energy redistribution corrections}
\label{response_appendix}

Energy redistribution effects can, in principle, modify the shape of energy-resolved pulse profiles by redistributing photons as described by the response matrix. To quantify the impact of redistribution, we compare PF spectra derived directly in count space with those obtained after unfolding the channel--phase matrices through the instrumental response. The unfolding procedure is performed using the \texttt{nDspec} framework following the approach described by \citet{houck2000astronomical}, which allows the construction of energy--phase matrices expressed in flux density units, accounting for the detector response.

\begin{figure}
	\centering
	\includegraphics[width=0.9\columnwidth]{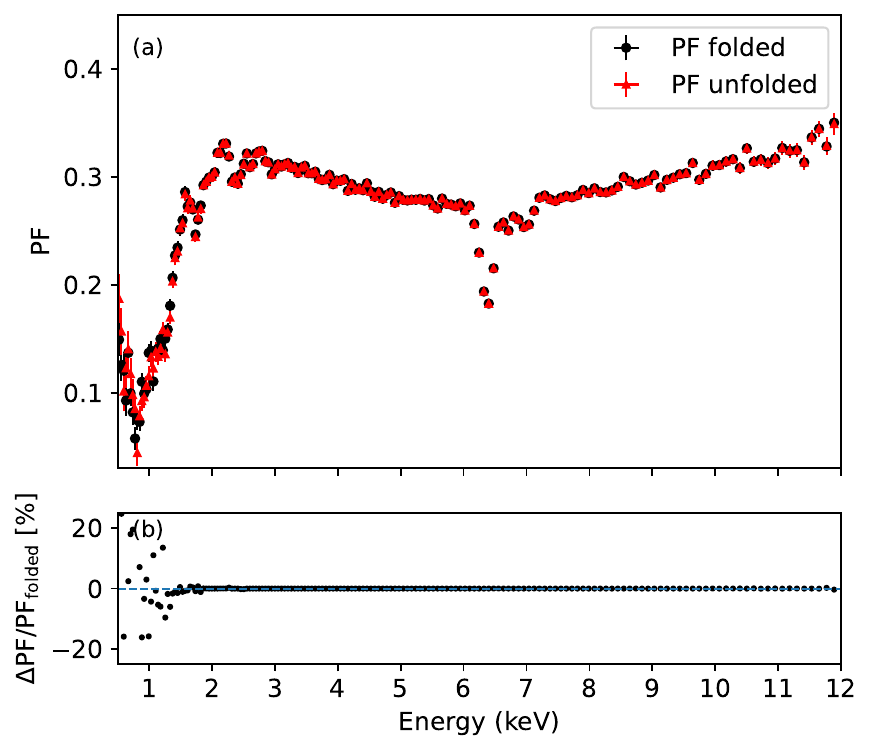}
	\caption{
	Effect of the response redistribution on the pulsed fraction spectrum derived from the \textit{XMM-Newton} data.
    Panel~(a) compares the pulsed fraction spectrum computed in count space with the one obtained in flux space
    after unfolding through the instrument response. Panel~(b) shows the corresponding relative difference
    (in percent) as a function of energy in the 0.5--12\,keV band.
	}
	\label{fig:pfs_response_xmm}
\end{figure}

For EPIC-pn, the comparison is shown in Fig. \ref{fig:pfs_response_xmm}, where the PF spectrum computed in count space is contrasted with that derived from unfolded energy--phase matrices in the 0.5--12\,keV band. The largest relative differences in PF occur at low energies. Above 1.5\,keV, the discrepancy stays below 10\%, and by 3\,keV it decreases to about 2\%. The stronger impact of energy redistribution at low energies is not unexpected, as this behavior has already been reported by \citet{haberl2002spectral}.

\begin{figure}
	\centering
	\includegraphics[width=0.9\columnwidth]{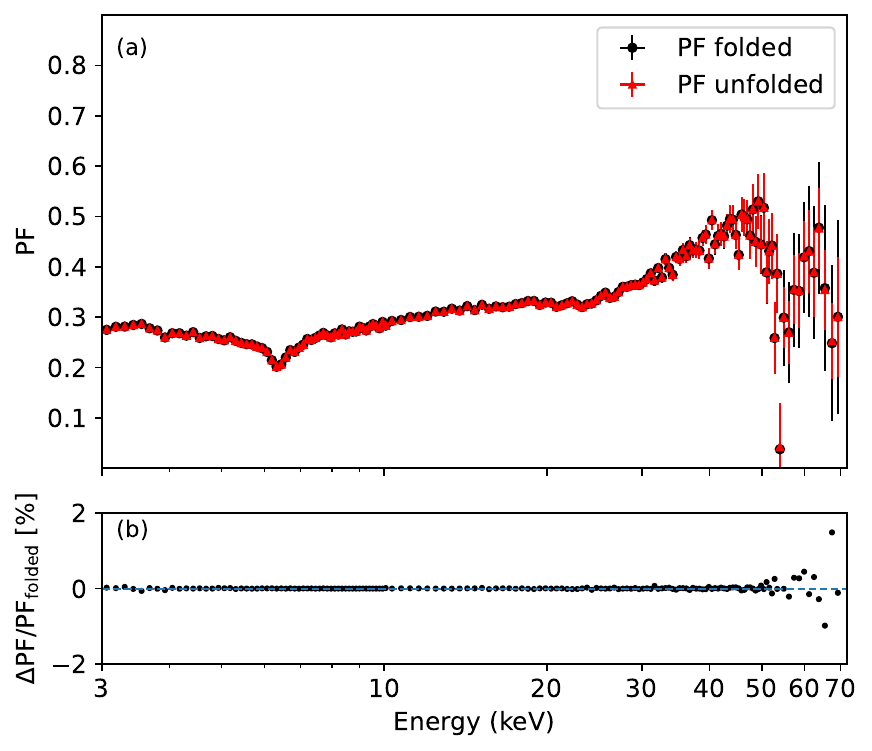}
	\caption{
    Effect of the response redistribution on the pulsed-fraction spectrum derived from the \textit{NuSTAR} data.
    Panel~(a) compares the pulsed fraction spectrum computed in count space with the one obtained in flux space
    after unfolding through the instrument response. Panel~(b) shows the corresponding relative difference
    (in percent) as a function of energy in the 3--70\,keV band.
	}
	\label{fig:pfs_response_nustar}
\end{figure}

An analogous test is performed for \textit{NuSTAR}, as shown in Fig.\ref{fig:pfs_response_nustar}, where PF spectra derived in count space and in flux space are compared over the 3--70\,keV band. The relative differences remain well below the percent level across the entire energy range (excluding the last energy bin). These results demonstrate that energy redistribution effects are negligible for \textit{NuSTAR}. Overall, these tests confirm that while there is a small effect at low energies in EPIC-pn, energy redistribution is largely negligible and does not significantly change the trends in either instrument.

\section{Impact of source region selection for pulse profiles extraction}
\label{regions_appendix}

\subsection{XMM--Newton}
\begin{figure}
	\centering
	\includegraphics[width=0.9\columnwidth]{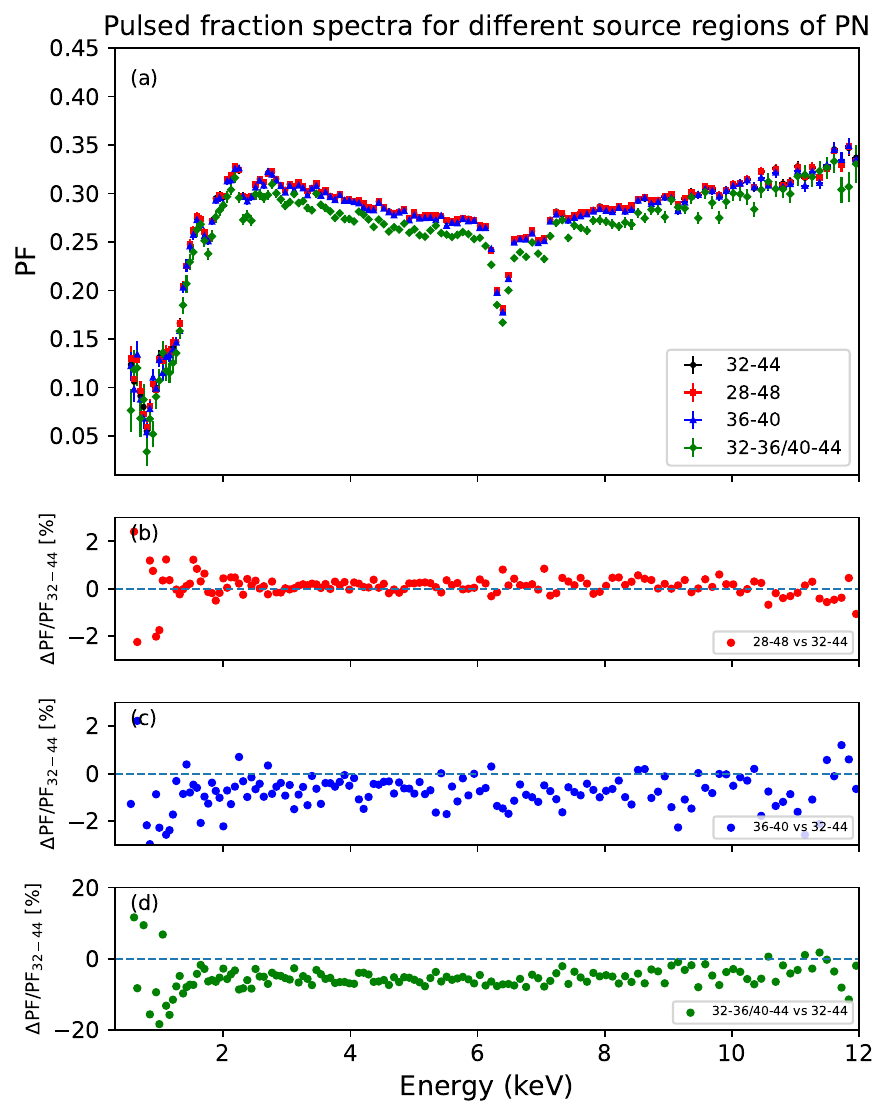}
	\caption{
		Pulsed fraction spectra obtained using four different source extraction regions for the EPIC-pn data.
		Panel~(a) shows the pulsed fraction as a function of energy in the 0.5--12\,keV band for each region.
		Panels~(b)--(d) show the relative difference (in percent) between the PF spectra obtained with the source region defined by RAWX in the 32--44 range and the other three test regions selected by RAWX in 36--40, 28--48, and (32--36 | 40--44) ranges, respectively.
	}
	\label{fig:pfs_regions_xmm}
\end{figure}

We tested the impact of the source region selection on the pulse profiles by considering four different extraction regions: (a) the same region adopted in \citet{Diez2023}, which excluded the central RAWX columns affected by pile-up (RAWX in [37--39]); (b) the same region including the piled-up columns; (c) a region defined only by the piled-up columns; and (d) a much larger region (RAWX in [28--48]) including the piled-up columns. The resulting PF spectra are shown in Fig.~\ref{fig:pfs_regions_xmm}. We find that the PF values are systematically lower when the piled-up columns are excluded. This indicates that removing the point-spread function core, and therefore a large fraction of the source counts, reduces the reconstructed pulse-profile amplitude. However, the overall energy dependence of the PF spectrum remains unchanged, indicating that the moderate level of pile-up present in the data does not introduce significant energy-dependent distortions in the pulse profiles. For the other three extraction regions, which include the central columns, the differences in PF remain below 2\% over the full energy range.

\subsection{NuSTAR}

We performed an analogous test for \textit{NuSTAR} by defining four different source regions, including one in which we removed the inner part of the source region (i.e.\ selecting an annular extraction region) to mimic the exclusion of the central EPIC-pn columns. In order to compare bin by bin, we rebinned all PF spectra to match that of the source region with the lowest SNR. We again found systematically lower PF values when the inner core of the point-spread function was excluded compared to regions that included it. Since pile-up is negligible in \textit{NuSTAR}, this indicates that the reduction of the measured PF is not caused by pile-up, but rather by the loss of source counts from the PSF core and the relatively larger contribution of the background in the remaining region. For the remaining extraction regions, the PF differences remain below 5\% across the whole energy range.

\begin{figure}
	\centering
	\includegraphics[width=0.9\columnwidth]{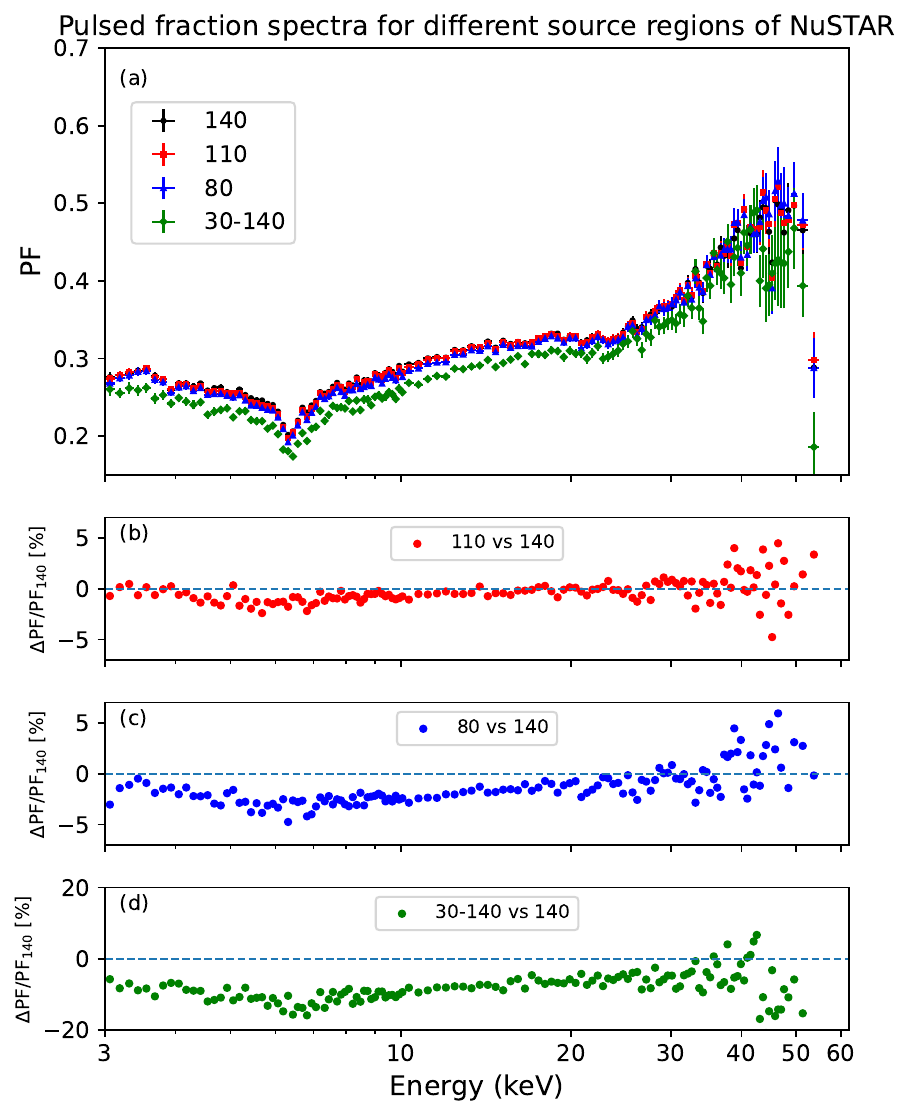}
	\caption{
		Pulsed fraction spectra obtained using four different extraction regions for the \textit{NuSTAR} data.
		Panel~(a) shows the pulsed fraction as a function of energy in the 3--70\,keV band for each region.
		Panels~(b)--(d) show the relative difference (in percent) between the pulsed fraction obtained with the circular
		source region of radius 140\arcsec{}, adopted in this work, and that derived with each of the other three regions.
		The label ``30--140'' refers to the annular region defined as the area between an inner circle of radius 30\arcsec{}
		and an outer circle of radius 140\arcsec{}.
	}
	\label{fig:pfs_regions_nustar}
\end{figure}

\end{document}